\documentclass[twocolumn,prb]{revtex4}
 \usepackage{epsfig}

\usepackage{dcolumn}
\usepackage{bm}

\begin{document}


\title{Gap Nodes and Time Reversal Symmetry
Breaking in Strontium Ruthenate}

\author{James F. Annett}
\affiliation{H. H. Wills Physics Laboratory, University of Bristol, Tyndall Ave,
BS8-1TL, UK.}

\author{G. Litak}
\affiliation{ Department of Mechanics, Technical University of Lublin,\\
Nadbystrzycka 36, 20-618 Lublin, Poland.}

\author{B. L. Gy\"orffy}
\affiliation{H. H. Wills Physics Laboratory, University of Bristol, Tyndall Ave,
 BS8-1TL, UK.}

\author{K. I. Wysoki\'nski}
\affiliation{Institute of Physics, M. Curie-Sk\l odowska University,\\
Radziszewskiego 10, 20-031 Lublin, Poland}
\date{\today}

\begin{abstract}
We study the superconducting state of Sr$_2$RuO$_4$ on the bases
of a phenomenological but orbital specific description of the
electron-electron attraction and a realistic quantitative account
of the electronic structure in the normal state. We found that a
simple model which features both `in plane' and `out of plane'
coupling with strengths $U_{\parallel}=40$meV and
$U_{\perp}=48$meV respectively reproduced the experimentally
observed power law behaviour of the low temperature specific heat
$C_v(T)$, superfluid density $n_s(T)$ and thermal conductivity in
quantitative detail. Moreover, it predicts that the quasi-particle
spectrum on the $\gamma$ -sheet is fully gaped and the
corresponding order parameter breaks the time reversal symmetry.
We have also investigated the stability of this model to inclusion
of further interaction constants in particular 
between orbitals contributing to 
 the $\gamma$ sheet of the
Fermi surface  and the $\alpha$ and $\beta$ sheets. We found that
the predictions of the model are robust under such changes.
Finally, we have incorporated a description of weak disorder into
the model and explored some of its consequences. For example we
demonstrated that the disorder has a more significant effect on
the $f$-wave component of the order parameter than on the $p$-wave
one.
\end{abstract}


\maketitle



\section{Introduction}

The symmetry of the order parameter
in superconducting Sr$_2$RuO$_4$
has been a subject of intense experimental and theoretical
interest in recent years\cite{maeno01,mackenzie03}.
It is probably the best candidate, currently, for
an odd-parity, spin triplet, superconductor which would be
a charged particle analogue of superfluid $^3$He.\cite{leggett75}
Although a number of other superconductors
are also possible spin-triplet superconductors (including
UPt$_3$, UGe$_2$, ZrZn$_2$,
and Bechgaard salts) strontium ruthenate is probably the one which is best
characterized experimentally. Samples can be grown which have exceptionally
long mean free paths,\cite{mackenzie00}  and  above $T_c$ the  normal state
is a Fermi liquid with a well understood Fermi surface\cite{bergemann00}.

Currently controversy exists over two key aspects
of the Sr$_2$RuO$_4$ pairing state.
Firstly, the gap function symmetry is still not known.
Rice and Sigrist\cite{rice95} suggested several
possible gap functions for Sr$_2$RuO$_4$ corresponding to
analogues of superfluid phases of $^3$He. Of these only the
analogue of the Anderson-Brinkman-Morel (ABM) state\cite{leggett75},
\begin{equation}
  \mathbf{d}(\mathbf{k}) \sim  (k_x + i k_y ) \hat{\mathbf{e}}_z ,
\label{eq:gapfunction}
\end{equation}
is consistent with the observations of a constant $a-b$ plane
Knight shift\cite{ishida98} and spin susceptibility\cite{duffy00}
below $T_c$. This state is also consistent with the $\mu$-SR
experiments which show spontaneous time reversal symmetry breaking
at $T_c$.\cite{luke98} However this gap function has no zeros on
the three cylindrical Fermi surface sheets\cite{bergemann00} of
Sr$_2$RuO$_4$,  in direct contradiction to several experiments
which indicate that the gap function has lines of zeros on the
Fermi surface\cite{nishizaki00,bonlade00,izawa00}. This
discrepancy is not easily resolved since a complete group
theoretic classifications of all symmetry distinct pairing states
in tetragonal
crystals\cite{volovik85,ozaki86,sigrist87,ozaki89,annett90,sigrist91}
 {\em do not include any states} which have
both spontaneous time reversal symmetry breaking at $T_c$ and
symmetry required line nodes on a cylindrical Fermi
surface.\cite{annett90} A number of {\it `f-wave'} gap functions
have been proposed\cite{won00,graf00,dahm00} for Sr$_2$RuO$_4$,
\begin{equation}
  \mathbf{d}(\mathbf{k}) \sim f(\mathbf{k}) \hat{\mathbf{e}}_z ,
\label{eq:gapfunctionone}
\end{equation}
where $f(\mathbf{k})$ is an $l=3$ spherical Harmonic function.
Such gap functions have constant $a-b$ plane Knight shift and may
have both time reversal symmetry breaking  and line nodes, however
in tetragonal symmetry crystals they  are always either of mixed
symmetry (requiring a double phase transition) or are in the same
symmetry class ($E_u$) as $l=1$ {\it `p-wave'} states which do not
have line nodes.  Such f-wave functions may be possible physically
(depending on the details of the actual pairing interaction), but
the line nodes are not required by the symmetry of the pairing
state.

The second controversy  about the Sr$_2$RuO$_4$ gap function
concerns the presence of three different Fermi surface sheets,
$\alpha$, $\beta$ and $\gamma$.  The {\em orbital dependent
superconductivity} model of Agterberg, Sigirst and
Rice\cite{agterberg97} envisioned a dominant gap on one part the
Fermi surface (originally $\alpha$, $\beta$), with the gap
function on the other band only arising from interband coupling
and hence being significantly smaller. This theory predicted that
weak impurity scattering would destroy the small gap on the
inactive sheet, and hence lead to a finite residual density of
states at zero energy. However the  experimental specific heat
data\cite{nishizaki00} shows that $C_V/T$ is zero at $T=0$, and
hence there is a finite order parameter on all sheets of the Fermi
surface. In a recent letter, Zhitomirsky and
Rice\cite{zhitomirsky01} have
 argued that the gap
function of superconducting strontium ruthenate can be described by
an effective, k-space, {\em interband-proximity effect}. In this model they
propose that the superconductivity is due to an attractive
interaction in the p-wave channel, which is acting almost entirely
on one sheet of the Fermi surface, the $\gamma$ sheet.
The other two Fermi surface sheets, $\alpha$ and $\beta$ are driven
to become superconducting because of a ``proximity effect''
or Josephson like coupling between the $\gamma$ and $\alpha$,$\beta$
bands.  This model has a number of features which are consistent
with the experimental facts, such as the presence of both
line-nodes in the gap function and spontaneous time reversal symmetry breaking
below $T_c$. Furthermore, if
the interband Josephson coupling energy is chosen to be
sufficiently large, then the energy gap at low
temperatures is moderately large on all the Fermi surface sheets
and there is no second peak
below $T_c$ in the specific
heat capacity.

In a recent paper we have proposed a quite general
semi-phenomenological methodology for studying the possible
superconducting states of Sr$_2$RuO$_4$. In this approach one
chooses, more or less systematically, orbital and position
dependent interaction constants to describe the electron-electron
attraction. The simplest useful model we have studied prominently
featured interlayer coupling\cite{annett01}. This model
characterizes the pairing interaction in terms of two
nearest-neighbor negative-$U$ Hubbard interactions, one,
$U_{\parallel}$ acts between Ru $d_{xy}$ in a single RuO$_2$
plane, while the second, $U_{\perp}$ acts between   Ru $d_{xz}$
$d_{yz}$  orbitals between planes. When these two parameters are
chosen so as to give a single phase transition temperature at the
observed $T_c$ of $1.5{\rm K}$ we find excellent agreement with
the measured specific heat, penetration depth and thermal
conductivity data. The gap function has both time reversal
symmetry breaking, but also horizontal lines of nodes  in the
planes $k_z=\pm \pi/c$ on the $\beta$ Fermi surface sheet. The
$\gamma$ sheet remains node-less, with a gap function of the form
${\bf d}({\bf k}) \sim (\sin{k_x}+i\sin{k_y})\hat{\bf e}_z$,
corresponding to the 2-d analogue of the $^3$He A-phase. The
predicted gap function is similar to that of Zhitomirsky and Rice
(ZR) \cite{zhitomirsky01}, but differs in that it is more or less
same size on all three Fermi surface sheets. Moreover, while ZR
rely on `proximity coupling'
 to avoid the double phase transition we exploit the freedom provided
by the experimental data and achieve
the same end by fixing both $U_{\parallel}$
and $U_{\perp}$ so that there is only one transition at the observed $T_c=1.5$K.

The purpose of this paper is to clarify a number of unresolved
questions concerning the {\em interlayer coupling} model.
Firstly we show in Section III that the results 
of the model are quite generic, and do not
 depend sensitively on the choice of the specific Hubbard model
parameters which we used in Ref.\cite{annett01}. Secondly we
examine the effects of weak disorder on the gap function (Section IV). 
We show
that weak disorder can suppress any f-wave components of the gap
function, while leaving the p-wave order parameter relatively
unchanged.
Finally in Section V we study the generalisation of our model by allowing for
a "bond proximity" interactions. Such symmetry mixing interactions have
 been proposed by  Zhitomirsky and Rice\cite{zhitomirsky01} as a mechanism
leading to the single superconducting transition temperature. It 
turns out that the mechanism operates   in the orbital
picture as well and we obtained the single superconducting transition
temperature, but different slope and jump of the specific heat.   

\section{Gap Symmetry and Pairing Basis Functions}

Let us begin by reviewing briefly the symmetry principles which
are used to classify different pairing symmetry states in
odd-parity superconductors.  We shall use these principles to
contrast the different pairing states that have been proposed
for strontium ruthenate.

On very general ground we expect that the phase transition
into the superconducting state is of second order, and so there
exists an order parameter, or set of order parameters,
$\eta_i({\bf r}), i=1,\dots n$.  For superconductors these
order parameters are complex,
transforming under the $U(1)$ gauge symmetry
 as $ \eta_i \rightarrow e^{i\theta} \eta_i$. Therefore the
 Ginzburg-Landau Free energy can always be expanded as
 \begin{eqnarray}
     F_s = F_n  &+ &\int d^3r \left(
     \frac{\hbar^2}{2m_{ijkl}}  \partial_i \eta^*_j({\bf r})
     \partial_k \eta_l({\bf r}) \right. \nonumber \\
     && + \alpha_{ij} \eta^*_i({\bf r})\eta_j({\bf r}) \nonumber \\
      && + \left.\beta_{ijkl}
      \eta^*_i({\bf r})\eta^*_j({\bf r})\eta_k({\bf r})\eta_k({\bf r})
       + \dots \right)
 \end{eqnarray}
where summation convention is implied for the indices $i$,$j$ etc,
and as usual $\partial_i \equiv \nabla_i -2ei A_i/\hbar$, with
${\bf A}$ the magnetic vector potential.

If the normal state above $T_c$ possesses a symmetry group $\mathcal{G}$,
then the order parameters $\eta_i$ can be grouped into terms
corresponding to the different irreducible representations
$\Gamma$ of $\mathcal{G}$, transforming under symmetry operations
as
\begin{equation}
      \eta_i^\Gamma  \rightarrow  R^\Gamma_{ij}(g) \eta_j^\Gamma
\end{equation}
where $ g \in \mathcal{G}$, and the matrices $R^\Gamma_{ij}(g)$
constitute the representation $\Gamma$ of the group $\mathcal{G}$.

The general theory of group representations implies that we can
choose a basis in which the matrix $\alpha_{ij}$ is block
diagonal, with each block corresponding to an irreducible
representation, $\Gamma$. In this basis the full Ginzburg-Landau
Free energy is of the form
\begin{eqnarray}
     F_s &=& F_n  + \int d^3r \left(  \sum_{\Gamma, \Gamma'}
     \frac{\hbar^2}{2m^{\Gamma \Gamma'}_{ijkl}}
     \partial_i \eta^{\Gamma*}_j({\bf r})
     \partial_k \eta^{\Gamma'}_l({\bf r})
     \right. \nonumber \\
     && + \sum_\Gamma\alpha^\Gamma_{ij} \eta^{\Gamma*}_i({\bf r})
     \eta^\Gamma_j({\bf r})   \nonumber \\
     &&  +  \left.\sum_{\Gamma \Gamma' \Gamma'' \Gamma'''}
       \beta^{\Gamma \Gamma' \Gamma'' \Gamma'''}_{ijkl}
       \eta^{\Gamma*}_i({\bf r})\eta^{\Gamma'*}_j({\bf r})
       \eta^{\Gamma''}_k({\bf r})\eta^{\Gamma'''}_l({\bf r})
        \right) . \nonumber \\
        \label{eqglfreeenergy}
\end{eqnarray}
The quadratic
term $ \alpha^\Gamma_{ij}$  involves only a single representation, $\Gamma$.
At $T_c$,
in general, only a single irreducible representation
will have a zero eigenvalue
of the block diagonal matrix $\alpha^\Gamma_{ij}$. Therefore only
the components of the order parameter $\eta_i^\Gamma$ corresponding
to that eigenvector will become non-zero just below $T_c$.

\begin{table}
\caption{\label{tableone} Irreducible representations of
even and odd  parity in a tetragonal crystal. The symbols $X$, $Y$ $Z$
represent any functions transforming as $x$, $y$ and $z$ under
crystal point group operations, while $I$ represents
any function which is invariant under all point group symmetries. }
\begin{ruledtabular}
\begin{tabular}{rl|rl}
Rep. &  symmetry  & Rep. &  symmetry    \\
 \hline
 A$_{1g}$ & $I$ &               A$_{1u}$ & $XYZ(X^2-Y^2)$  \\
  A$_{2g}$ & $XY(X^2-Y^2)$ &    A$_{2u}$ & $Z$  \\
  B$_{1g}$ & $X^2-Y^2$  &     B$_{1u}$ & $XYZ$  \\
 B$_{2g}$ & XY &  B$_{2u}$ & $Z(X^2-Y^2)$ \\
  E$_g$  & $\{XZ,YZ\}$ &  E$_u$  & $\{X,Y\}$ \\
\end{tabular}
\end{ruledtabular}
\end{table}

\begin{table}
   \caption{\label{tabletwo} Products of the irreducible representations
   of $D_{4h}$ point group symmetry }
 \begin{ruledtabular}
 \begin{tabular}{l|ccccc}
 $\otimes$ & A$_1$  &   A$_2$ &  B$_1$  &   B$_2$ &  E\\
 \hline
 A$_{1}$  & A$_{1}$  & A$_2$ &  B$_1$  &   B$_2$ &  E\\
  A$_{2}$ & A$_{2}$  & A$_1$ &  B$_2$  &   B$_1$ &  E\\
  B$_{1}$ & B$_{1}$  &B$_2$ &  A$_1$  &   A$_2$ &  E \\
 B$_{2}$ &  B$_{2}$  &B$_1$ &  A$_2$  &   A$_1$ &  E\\
  E  &  E & E & E & E & A$_1$ $\oplus$ A$_2$ $\oplus$ B$_1$ $\oplus$ B$_2$\\
 \end{tabular}
\end{ruledtabular}
  \end{table}

Now let us apply these very general principles to the specific case
of  spin-triplet pairing in
Sr$_2$RuO$_4$. This is a body-centred tetragonal crystal with inversion
symmetry.   The relevant crystal group is
$D_{4h}$, and
Table \ref{tableone} shows its
irreducible representations. For each representation
its symmetry is denoted by a typical function, where
the symbols $X$, $Y$, $Z$ represent any
functions which transform as $x$, $y$ and $z$ under the point group
operations, and $I$ means any function which is invariant under
 all point group operations.   The representations
 A$_{1g}$ \dots E$_g$ have even parity, while A$_{1u}$ \dots E$_u$
 have odd parity.   Table \ref{tabletwo}
shows the multiplication table for the irreducible
representations,  i.e. how  direct products
of representation matrices $\Gamma \otimes \Gamma'$ decompose
into a sums of block diagonal matrices $\Gamma_1 \oplus \Gamma_2 \oplus \dots$.

An immediate consequence of the multiplication table
~\ref{tabletwo} is that in tetragonal crystals the
order parameter is either of a single representation $\Gamma$ only,
or there are two or more distinct thermodynamic phase transitions.
This is because to quadratic (or higher) order in
the Ginzburg-Landau free energy  there are no symmetry allowed
coupling terms of the form
$$
       \beta^{\Gamma' \Gamma \Gamma \Gamma}_{ijkl}
       \eta^{\Gamma'*}_i({\bf r})\eta^{\Gamma}_j({\bf r})
       \eta^{\Gamma}_k({\bf r})\eta^{\Gamma}_l({\bf r})
$$
in Eq.~\ref{eqglfreeenergy}.  The proof\cite{annett90} is
simply that $\Gamma' \otimes \Gamma \otimes \Gamma \otimes \Gamma$
never contains the identity representation $A_{1g}$, and hence
such terms are not allowed as quartic invariants of the Free energy
(or at higher order). In the absence of such terms the free
energy functional is always of at least quadratic
order in the subdominant order parameter
$\eta^{\Gamma'}_i({\bf r})$, and hence these subdominant components
can only become non-zero in a separate phase transition
below $T_c$.

\begin{table}
   \caption{\label{tablethree}
  Basis functions $\gamma_i^\Gamma({\bf k})$
  for the  odd parity irreducible representations
  of body-centred tetragonal crystals.}
 \begin{ruledtabular}
 \begin{tabular}{rll}
 Rep.  & in-plane & inter-plane \\
 \hline
 A$_{1u}$  &  -  &  -  \\
 A$_{2u}$  & -  &  $ \cos{\frac{k_x}{2}} \cos{\frac{k_y}{2}}
  \sin{\frac{k_zc}{2}}$\\
  B$_{1u}$  &- &
$ \sin{\frac{k_x}{2}} \sin{\frac{k_y}{2}}  \sin{\frac{k_zc}{2}}$ \\
 B$_{2u}$  & -  & - \\
 E$_u$   & $\sin{k_x}$   &
  $ \sin{\frac{k_x}{2}} \cos{\frac{k_y}{2}}  \cos{\frac{k_zc}{2}}$\\
   &     $\sin{k_y} $  &
      $ \cos{\frac{k_x}{2}} \sin{\frac{k_y}{2}}  \cos{\frac{k_zc}{2}}$\\
 \end{tabular}
\end{ruledtabular}
  \end{table}
\begin{figure}
\centerline{\epsfig{file=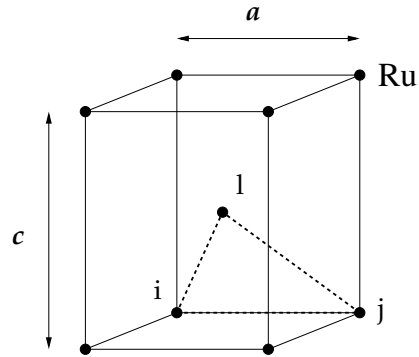,width=5.0cm,angle=0}}
 \caption{\label{fig_one}
  Body-centred tetragonal lattice, showing the nearest neighbour
 pairs in-plane, and  between planes.}
 \end{figure}

Using these irreducible representations we can  expand
the BCS gap function in terms of functions
of each separate symmetry class.
For odd parity pairing states we can represent the BCS gap function
by a vector ${\bf d}({\bf k})$ or a symmetric $2 \times 2$
complex matrix
\begin{equation}
   \left(  \begin{array}{cc}
     \Delta_{\uparrow\uparrow}({\bf k}) &
     \Delta_{\uparrow\downarrow}({\bf k}) \\
     \Delta_{\uparrow\downarrow}({\bf k}) &
     \Delta_{\downarrow\downarrow}({\bf k})
  \end{array}
\right) =
\left(  \begin{array}{cc}
     id_y({\bf k})-d_x({\bf k}) & d_z({\bf k}) \\
    d_z({\bf k})  &  d_x({\bf k})+ id_y({\bf k})
     \end{array}  \right)
\end{equation}
where $\Delta_{\uparrow\downarrow}({\bf k})=
 \Delta_{\downarrow\uparrow}({\bf k})$
and $\Delta_{\sigma\sigma'}({\bf k}) = - \Delta_{\sigma\sigma'}(-{\bf k}) $.
For each irreducible representation we can choose a complete
set of orthonormal basis functions in the
Brillouin zone, $\gamma^\Gamma_i({\bf k})$.
Expanding the gap function  in terms of these functions we have
\begin{equation}
   \Delta_{\sigma\sigma'}({\bf k}) = \sum_i \Delta^\Gamma_{i\sigma\sigma'}
   \gamma^\Gamma_i({\bf k}).
 \end{equation}
The expansion coefficients essentially provide the set of
order parameters in Eq.\ref{eqglfreeenergy}.
The basis functions must be periodic in reciprocal space,
$\gamma^\Gamma_i({\bf k})=\gamma^\Gamma_i({\bf k}+{\bf G})$,
or equivalently, they must obey periodic boundary conditions in the 1st
Brillouin zone.
They can be chosen, most naturally, in terms of their
real-space Fourier transforms, which correspond to lattice
sums of the real-space Bravais lattice.
For a body-centred tetragonal crystal, such as Sr$_2$RuO$_4$
shown in Fig.~\ref{fig_one}, the  leading basis functions
correspond to the four nearest-neighbour in-plane
lattice vectors, ${\bf R}= \pm a \hat{\bf e}_x$
and   ${\bf R}=\pm a \hat{\bf e}_y$,
giving two odd parity basis functions: $\sin{k_xa}$ and $\sin{k_ya}$.
The eight body-centred lattice vectors $
{\bf R}= \pm \frac{a}{2} \hat{\bf e}_x
 \pm \frac{a}{2} \hat{\bf e}_y \pm \frac{c}{2}\hat{\bf e}_z$
lead to the four odd-parity basis functions shown in the last column of
Table \ref{tablethree} (where for simplicity we have chosen units
of length such that $a=1$).
In the models which we investigate
in the remainder of this paper, we shall assume that these
basis functions, Table \ref{tablethree}, are sufficient
to describe the gap function. Physically this corresponds to the assumption
that the paring interaction $V_{\sigma \sigma'}({\bf r},{\bf r}')$
is short ranged in real-space.

Considering Table~\ref{tableone}
we can see that in Sr$_2$RuO$_4$ ``p-wave'' pairing states can correspond
to either the A$_{2u}$, (or $p_z$) representation
or the doubly degenerate E$_u$ representation ($p_x$, $p_y$).
The only symmetry distinct ``f-wave'' pairing states are
the B$_{1u}$  and B$_{2u}$ representations, corresponding
to $f_{xyz}$  and $f_{(x^2-y^2)z}$ type symmetries.
Neither of these states can be used in the case of a
two-dimensional single-plane
model of Sr$_2$RuO$_4$, since they both become zero in the plane
$k_z=0$.   It is also interesting to note that
in Table~\ref{tablethree} there are no
basis functions of $A_{1u}$ or $B_{2u}$ symmetry. Pairing in these
channels would require long range interactions
extending to at least the
inter-plane second nearest neighbors.

 In the light of these symmetry principles let us comment on a number of
 the possible gap functions which have been proposed for
 Sr$_2$RuO$_4$.  Among the five states described by
Rice and Sigrist\cite{rice95} the only one consistent with the
Knight shift
experiments is\cite{annett90}
\begin{equation}
    {\bf d}({\bf k}) = (\sin{k_x}+i\sin{k_y}) \hat{\bf e}_z
 \end{equation}
 belonging to the
 E$_u$ representation of Table~\ref{tablethree}. It breaks time
 reversal symmetry, consistent with the $\mu$-SR experiments
 of Luke {\it et al.}\cite{luke98}, and leads to
 a spin susceptibility  which is constant below $T_c$
 for fields in the $a-b$ plane, consistent with Knight shift\cite{ishida98}
 and neutron scattering experiments
 \cite{duffy00}.  However it
 has no gap nodes
 on a Fermi surface of cylindrical  topology, such as the
 $\alpha$, $\beta$  and $\gamma$ sheets of Sr$_2$RuO$_4$, and therefore
 is inconsistent with the heat capacity\cite{nishizaki00}
 penetration depth\cite{bonlade00} and thermal conductivity
 experiments\cite{izawa00}.

 On the other hand the f-wave gap function proposed by
 Won and Maki\cite{won00}
\begin{equation}
   {\bf d}({\bf k}) \sim  k_z( k_x \pm i k_y)^2 \hat{\bf e}_z
   \label{wonmaki}
 \end{equation}
 has both line nodes and broken time reversal symmetry below
 $T_c$. However from the symmetry analysis above, it is clear
that this
 does not correspond to a single irreducible representation of the symmetry
 group. It is a sum of the function $ k_z( k_x^2-k_y^2)$, belonging to
 B$_{2u}$ and  $ k_xk_yk_z$ belonging to B$_{1u}$.  Although
 they would be degenerate in a system with cylindrical symmetry,
in a tetragonal crystal they will be non-degenerate
 and hence have different $T_c$s. The B$_{1u}$, B$_{2u}$
 states individually possess time reversal symmetry. Therefore
 with this order parameter we would expect to find a specific heat anomaly with
 two transitions, and time reversal symmetry breaking would only
 occur at temperatures below the lower transition.

 The f-wave order parameter proposed by Graf and Balatsky\cite{graf00},
\begin{equation}
   {\bf d}({\bf k}) \sim  k_x k_y ( k_x + i k_y) \hat{\bf e}_z
   \label{grafbalatsky}
 \end{equation}
 is in the same symmetry class as E$_u$, since ${\rm B}_{2} \otimes {\rm E} =
 {\rm E}$ in Table~\ref{tabletwo}.
  Therefore in the sense of pure symmetry arguments the
 gap nodes in  planes $k_x=0$ and $k_y=0$ are ``accidental''.
 Such a gap function is certainly valid, but the nodes are present for
 reasons connected with the specific microscopic pairing interaction
 employed, and not required by symmetry alone.
 This comment also applies to the ${\rm B}_{1} \otimes {\rm E} $
 f-wave state
\begin{equation}
   {\bf d}({\bf k}) \sim  (k_x^2 - k_y^2) ( k_x + i k_y) \hat{\bf e}_z
   \label{eq:dahmwonmaki}
 \end{equation}
 discussed by Dahm, Won and Maki,\cite{dahm00} and Eremin {\it et
 al.}\cite{eremin01,manske02}.

 The full group theoretic classification
 in tetragonal
 crystals\cite{volovik85,ozaki86,sigrist87,ozaki89,annett90,sigrist91}
 and the above analysis does not show {\em a single pairing state}
 with both symmetry required lines of nodes and
 spontaneously broken time reversal symmetry below $T_c$.
 Therefore, if we accept both the $\mu$-SR and low
 temperature thermodynamic and transport measurements, then we must
 consider states which have lines of nodes for
 specific microscopic reasons, rather then for pure symmetry reasons.

 In the remainder of this paper we shall focus on the specific
 model which we proposed in a previous paper\cite{annett01},
 in which the lines of nodes appear in the plane $k_z = \pm \pi/c$,
 derived from the pair of inter-plane basis functions
 of ${\rm E}_u$:
 $$\sin{\frac{k_x}{2}}\cos{\frac{k_y}{2}}\cos{\frac{k_zc}{2}},
 \hspace*{0.5cm}
 \cos{\frac{k_x}{2}}\sin{\frac{k_y}{2}}\cos{\frac{k_zc}{2}}
 $$
 from Table\ref{tablethree}, as originally suggested by Hasegawa
 {\it et al.}\cite{hasegawa00}.

\section{Interlayer Coupling Hamiltonian}

\begin{figure}
\centerline{\epsfig{file=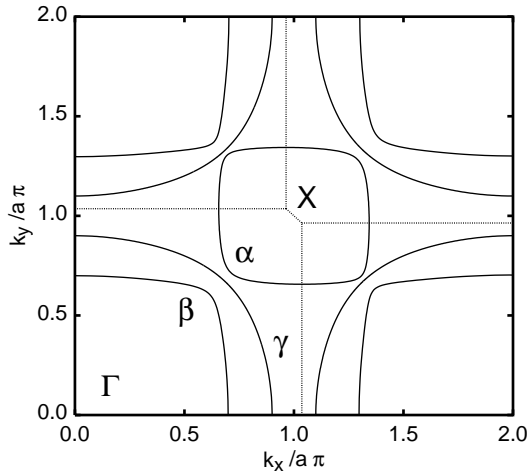,width=7.5cm,angle=-90}}
 \caption{\label{fig_two}
 The Fermi surface of Sr$_2$RuO$_4$
in the plane $k_z=0$, obtained by fitting the de Hass data of
Bergman {\it et al.}\cite{bergemann00}. Note that the alpha Fermi
surface sheet has only two-fold symmetry, because of the shape of
the Brillouin zone boundary.}
 \end{figure}

Since the underlying microscopic mechanism for superconductivity
in Sr$_2$RuO$_4$ is not known we choose to adopt a
phenomenological approach to the pairing mechanism. We first make
an accurate tight binding fit to the experimentally determined
Fermi surface\cite{mackenzie96,bergemann00} and then introduce
model attractive interactions between the different orbitals
centered on different sites. We can investigate different
`scenarios' depending upon which model interactions are assumed to
dominate. Frequently, when these pairing interaction parameters
are chosen to reproduce the experimental $T_c$, there is  no
freedom to adjust the parameters further. Once the parameters have
been selected, then a number of different experimental quantities
can be calculated independently and compared to experiment. The
goal is to find one specific paring scenario which agrees with all
of the experimental observations. If this can be achieved then one
has found an effective Hamiltonian for the pairing, which can be
interpreted physically. This effective pairing Hamiltonian can
then be used to guide the search for the true microscopic
Hamiltonian. This methodology has proved very useful in cuprate
superconductivity \cite{szotek} and here we shall deploy it to
study Sr$_2$RuO$_4$.

The effective pairing Hamiltonian we consider is
a simple multi-band
attractive $U$ Hubbard model:
\begin{eqnarray}
  \hat{H}& =& \sum_{ijmm',\sigma}
\left( (\varepsilon_m  - \mu)\delta_{ij}\delta_{mm'}
 - t_{mm'}(ij) \right) \hat{c}^+_{im\sigma}\hat{c}_{jm'\sigma} \nonumber \\
&& - \frac{1}{2} \sum_{ijmm'\sigma\sigma'} U_{mm'}^{\sigma\sigma'}(ij)
 \hat{n}_{im\sigma}\hat{n}_{jm'\sigma'} \label{hubbard}
\end{eqnarray}
 where $m$ and
$m'$ refer to the three Ruthenium $t_{2g}$ orbitals $a=xz$,
$b = yz$ and $c = xy$ and  $i$ and $j$ label the sites
of a body centered tetragonal lattice.

The hopping integrals $t_{mm'}(ij)$ and site energies
$\varepsilon_m$ were fitted to reproduce the experimentally
determined Fermi Surface \cite{mackenzie96,bergemann00}. The
nearest neighbour in-plane hopping integrals along ${\bf
R}=\hat{\bf e}_x$, where the {\it ab} plane lattice constant
is taken to be 1, are constrained by the orbital symmetry to have
the following form
\begin{equation}
   [t_{mm'}]  = \left( \begin{array}{ccc}
      t_{ax} &  0 & 0 \\
      0 &  t_{bx} & 0 \\
      0 & 0 & t
   \end{array}\right)
\end{equation}
(and similarly for ${\bf R}=\hat{\bf e}_y$ taking into account
sign changes due to orbital symmetries). The next
nearest neighbour in-plane hopping integrals along $\hat{\bf e}_x +
\hat{\bf e}_y$ were assumed to be of the form
\begin{equation}
   [t_{mm'}]  = \left( \begin{array}{ccc}
      0 &  t_{ab} &  0 \\
      t_{ab} &  0 & 0 \\
      0 & 0 & t'
   \end{array}\right).
\end{equation}
The parameter $t'$ controls the shape of the $\gamma$-band Fermi
surface, while the parameter $t_{ab}$ determines the hybridization
between the $a$ and $b$ orbitals and hence the shape of the
$\alpha$ and $\beta$ Fermi surfaces. The $c-$-axis magnetic field
de Hass van Alphen data\cite{mackenzie96} gives the areas and
cyclotron masses of the three Fermi surface sheets, and these six
numbers can be fit exactly with
 $t=0.08162{\rm eV}$, $t'=-0.45t$, $t_{ax}=
1.34t$, $t_{bx}=0.06t_{ax}$, $t_{ab}=0.08t_{ax}$,
and the on-site energies were $\varepsilon_c=-1.615t$
and $\varepsilon_a=\varepsilon_b=-1.062t_{ax}$.

To obtain a three dimensional Fermi surface we assumed that the
dominant inter-plane hopping is along the body-centre vector ${\bf
R}=\frac{1}{2}(\hat{\bf e}_x +\hat{\bf e}_y + c \hat{\bf e}_z)$
and has the form
\begin{equation}
   [t_{mm'}]  = \left( \begin{array}{ccc}
      t_\perp & t_{hyb}  &  t_{hyb} \\
      t_{hyb} &  t_\perp & t_{hyb} \\
      t_{hyb} & t_{hyb} &  0
   \end{array}\right)
\end{equation}
and similarly for ${\bf R}=\frac{1}{2}(\pm \hat{\bf e}_x \pm
\hat{\bf e}_y \pm c \hat{\bf e}_z)$ with appropriate sign
changes. The parameter $t_{hyb}$ is the only term in the
Hamiltonian which mixes the $c$ orbitals with $a$ and $b$. With
only these two parameters it is not possible to fit exactly the
full three dimensional Fermi surface cylinder corrugations
determined by Bergemann {\it et al.}\cite{bergemann00}, but the
parameters $t_{hyb}=0.12t_{ab}$, $t_\perp=-0.03t_{ab}$ give a
reasonable agreement for the dominant experimental corrugations.
Fig.~\ref{fig_two} shows the fitted Fermi surface in the plane
$k_z=0$ in the extended zone-scheme. Note that the $\alpha$ sheet
has only two-fold symmetry, due to its position centred on the
Brillouin zone boundary at $X$.

The set of interaction
constants $U_{mm'}^{\sigma\sigma'}(ij)$ describe attraction
between electrons on nearest neighbour sites
with spins $\sigma$ and $\sigma'$
and in orbitals $m$ and $m'$.
Thus our actual calculations consists of solving, self-consistently,
the following Bogoliubov-de Gennes equation:
\begin{equation}
 \sum_{jm'\sigma'} \left(\begin{array}{c}
 E^\nu - H_{mm'}(ij)  ~ ~  ~
 \Delta^{\sigma\sigma'}_{m,m'} (ij)\\
 \Delta^{* \sigma\sigma'}_{mm'}(ij) ~ ~ ~ ~
 E^\nu +  H_{mm'}(ij)
\end{array}\right)
\left(\begin{array}{ll}
 u^\nu_{j m'\sigma'}\\
v^\nu_{jm'\sigma'}\end{array}\right)=0\,, \label{bogoliubov}
\end{equation}
where  $ H_{mm'}(ij) $
is the normal spin independent part of the Hamiltonian, and
the $\Delta^{\sigma\sigma'}_{mm'}(ij)$ is
self consistently given
in terms of the pairing amplitude, or order parameter,
$\chi_{mm'}^{\sigma\sigma'}(ij)$,
\begin{equation}
 \Delta^{\sigma\sigma'}_{mm'}(ij) = U_{mm'}^{\sigma\sigma'}(ij)
\chi_{mm'}^{\sigma\sigma'}(ij)\,. \label{deltas}
\end{equation}
defined by the usual relation
\begin{equation}
\chi_{mm'}^{\sigma\sigma'}(ij) =
\sum_{\nu} u^\nu_{im\sigma}v^{\nu *}_{jm'\sigma'}
(1 - 2f(E^\nu))\,,
\label{eq18}
\end{equation}
where $\nu$ enumerates the solutions of Eq.~\ref{bogoliubov}.

We solved the above system of Bogoliubov de Gennes equations
including all three bands and the three dimensional tight-binding
Fermi surface. We considered  a large number of different
scenarios for the interaction constants. First we assumed that the
pairing interaction $U_{mm'}^{\sigma\sigma'}(ij)$ for nearest
neighbours in plane is only acting for the $c$  ($d_{xy}$) Ru
orbitals. In this case both a d-wave ($d_{x^2-y^2}$) pairing state
and p-wave ($(k_x+ik_y) \hat{\bf e}_z)$ states are possible. The
d-wave state has line nodes, but would not be consistent with the
experiments showing constant Knight shift and time reversal
symmetry breaking below $T_c$. Therefore we discard such solutions
here, and only concentrate on the odd-parity spin triplet
solutions. The motivation is not to explain the microscopic
pairing mechanism, but to model pairing state produced by various
types of effective attractive interactions. These attractive
interactions may arise from, for instance, ferromagnetic spin
fluctuations\cite{rice95,miyake,eremin01,manske02}, which can
favour spin triplet pairing compared to the d-wave solutions.
However, their origin may be more complicated, for example a
combined electron-phonon and spin fluctuation mechanism.

With only the nearest neighbor in-plane interactions the set of
possible odd-parity, spin triplet, solutions that we found never
includes any possible state with nodes of the gap. Therefore we
extended the model to include inter-plane interactions.  Using two
interactions, a nearest neighbor in-plane interaction,
 ($i-j$ in Fig.~\ref{fig_one}), and a nearest
neighbor inter-plane interaction,  ($i-l$ in Fig.~\ref{fig_one})
which fulfill the tetragonal symmetry, we have the two types of
basis functions for the gap equation given in
Table~\ref{tablethree}. Then we have the possibility of horizontal
line nodes in the gap arising from the zeros of $\cos{(k_zc/2)}$
at $k_z=\pi/c$ on a cylindrical Fermi surface\cite{hasegawa00}.

Because the pairing interactions
 $U^{\sigma\sigma'}_{mm'}(ij)$ were assumed to act
only for nearest neighbor sites in or out of plane, the pairing
potential $ \Delta^{\sigma\sigma'}_{mm'}(ij)$ is also restricted
to nearest neighbors. We further focus on only odd parity (spin
triplet) pairing states for which the vector ${\bf d} \sim
(0,0,d^z)$, i.e. $ \Delta^{\uparrow\downarrow}_{mm'}(ij)=
  \Delta^{\downarrow\uparrow}_{mm'}(ij)$, and
$ \Delta^{\uparrow\uparrow}_{mm'}(ij)=
\Delta^{\downarrow\downarrow}_{mm'}(ij)=0 $. Therefore in general
we have the following non-zero order parameters  (i) for in plane
bonds: $\Delta_{mm'}^\parallel(\hat{\bf e}_x)$,
$\Delta_{mm'}^\parallel(\hat{\bf e}_y)$, and (ii) for inter-plane
bonds: $\Delta_{mm'}^\perp({\bf R}_{ij})$ for ${\bf R}_{ij}=(\pm
a/2, \pm a/2, \pm c/2)$.

Taking the lattice Fourier transform of Eq.~\ref{deltas}
the corresponding pairing potentials in k-space
have the general form (suppressing the spin indices for clarity):
\begin{eqnarray}
 \Delta_{mm'}({\bf k})  &=&  \Delta_{mm'}^{\parallel p_x}  \sin{k_x} +
 \Delta_{mm'}^{\parallel p_y}  \sin{k_y} \nonumber \\
 && + \Delta^{\perp p_x}_{mm'} \sin{\frac{k_x}{2}}
\cos{\frac{k_y}{2}}  \cos{\frac{k_zc}{2}}\nonumber \\
&& +\Delta^{\perp p_y}_{mm'} \sin{\frac{k_y}{2}}
\cos{\frac{k_x}{2}} \cos{\frac{k_zc}{2}} \nonumber \\
 & & + \Delta^{\perp p_z}_{mm'} \sin{\frac{k_zc}{2}}
 \cos{\frac{k_x}{2}} \cos{\frac{k_y}{2}}
\nonumber \\
&&+\Delta^{\perp f}_{mm'}  \sin{\frac{k_x}{2}} \sin{\frac{k_y}{2}}
 \sin{\frac{k_zc}{2} }
..   \label{eq20}
 \end{eqnarray}
Note that beyond the usual p-wave
symmetry of the $\sin{k_x}$ and $\sin{k_y}$ type for the $c$
orbitals, we include all three additional p-wave symmetries of the
$\sin{k/2}$ type which are induced by the effective attractive
interactions between carriers on the neighboring out-of-plane Ru
orbitals. These interactions are also responsible for the f-wave
symmetry order parameters, $\Delta^{\perp f}_{mm'}$, transforming
as $B_{1u}$ in Table 1. This latter is symmetry distinct from all
p-wave order parameters in a tetragonal crystal, unlike the other
f-wave states
 discussed in the
introduction\cite{graf00,won00,dahm00,eremin01}. The $p_z$ order
parameters $\Delta^{\perp p_z}_{mm'}$ are of $A_{2u}$ symmetry. In
contrast the pairs $\Delta^{\perp p_x}_{mm'},\Delta^{\perp
p_y}_{mm'}$ are of the same $E_u$ {\it `p-wave'} symmetry as
$\Delta_{mm'}^{\parallel p_x},\Delta_{mm'}^{\parallel p_y}$. In
general, the order parameters in each distinct irreducible
representations have different transition temperatures, as
expected from Eq. \ref{eqglfreeenergy}.

In a recent paper\cite{annett01} we chose a particularly simple
set of attractive pairing interactions
 $U^{\sigma\sigma'}_{mm'}(ij)$.
For in-plane nearest neighbours we assumed that the pairing
interaction is only acting for the $c$ ($d_{xy}$) Ru orbitals only
\begin{equation}
U_{\parallel mm'}= \left( \begin{array}{ccc} 0 & 0 & 0 \\
0 & 0 & 0 \\
0 & 0 & U_{\parallel}
   \end{array} \right), {\rm~~where~~} U_{\parallel}=0.494t,
\label{eq21} .
\end{equation}
On the other hand,  given that the ruthenium $a$ and $b$ orbitals
($d_{xz}, d_{yz}$) are oriented perpendicularly to the planes we
choose to introduce the inter-plane interaction only for these
orbitals,
\begin{equation}
U_{\perp mm'}= \left( \begin{array}{ccc} U_{\perp} & U_{\perp} & 0 \\
U_{\perp} & U_{\perp} & 0 \\
0 & 0 & 0
   \end{array} \right), {\rm~~~where~~} U_{\perp}=0.590t.
\label{eq22}
\end{equation}
Therefore we have, as a minimal set,
 only two coupling constants $U_\parallel$ and
$U_\perp$ describing these two physically different interactions.

As discussed earlier our strategy is to adjust these
phenomenological parameters in order to obtain one transition at
the experimentally determined $T_c$. Thus, beyond fitting $T_c$,
there are no further adjustable parameters, and one can compare
directly the calculated physical properties of the superconducting
states to those  experimentally observed. Consequently, if
 one obtains a good overall agreement one can say
that one has empirically determined the form of the pairing
interaction in a physically transparent manner. Evidently such
conclusion is the principle aim of the calculations.

As we have shown in Ref.\onlinecite{annett01}, this two parameter
scenario gives an excellent agreement with the experimental
specific heat\cite{nishizaki00}, superfluid
density\cite{bonlade00} and thermal conductivity\cite{izawa00}. We
chose the constants $U_\parallel$ and $U_\perp$, so that there is
a single phase transition at $T_c=1.5{\rm K}$, corresponding to
the values given in Eqs.~\ref{eq21} and \ref{eq22}. Below $T_c$
 the order
parameters have the symmetries $\Delta_{cc}^{\parallel
p_y}=i\Delta_{cc}^{\parallel p_x}$, $\Delta_{bb}^{\perp
p_y}=i\Delta_{aa}^{\perp p_x}$ as expected for an $E_u$  pairing
symmetry\cite{agterberg97} $(k_x+ik_y)\hat{\bf e}_z$ corresponding
to the same time reversal broken pairing state as $^3He-A$. We
also found that a much lower temperatures, additional transitions
occurred where the $f-$wave and $p_z$ order parameters become
non-zero. The gap function has line nodes on the Fermi surface, in
agreement with experiment, only when the f-wave component is zero.
Arguing that the f-wave component would be suppressed by
impurities, we showed that with the f-wave component removed, one
obtains excellent agreement between the calculated and
experimental specific heat, penetration depth and thermal
conductivity. We show, in Sec. IV below, that this removal of the
{\it f-wave} component is justified by the presence of weak
disorder.
 
\begin{figure}
\centerline{\epsfig{file=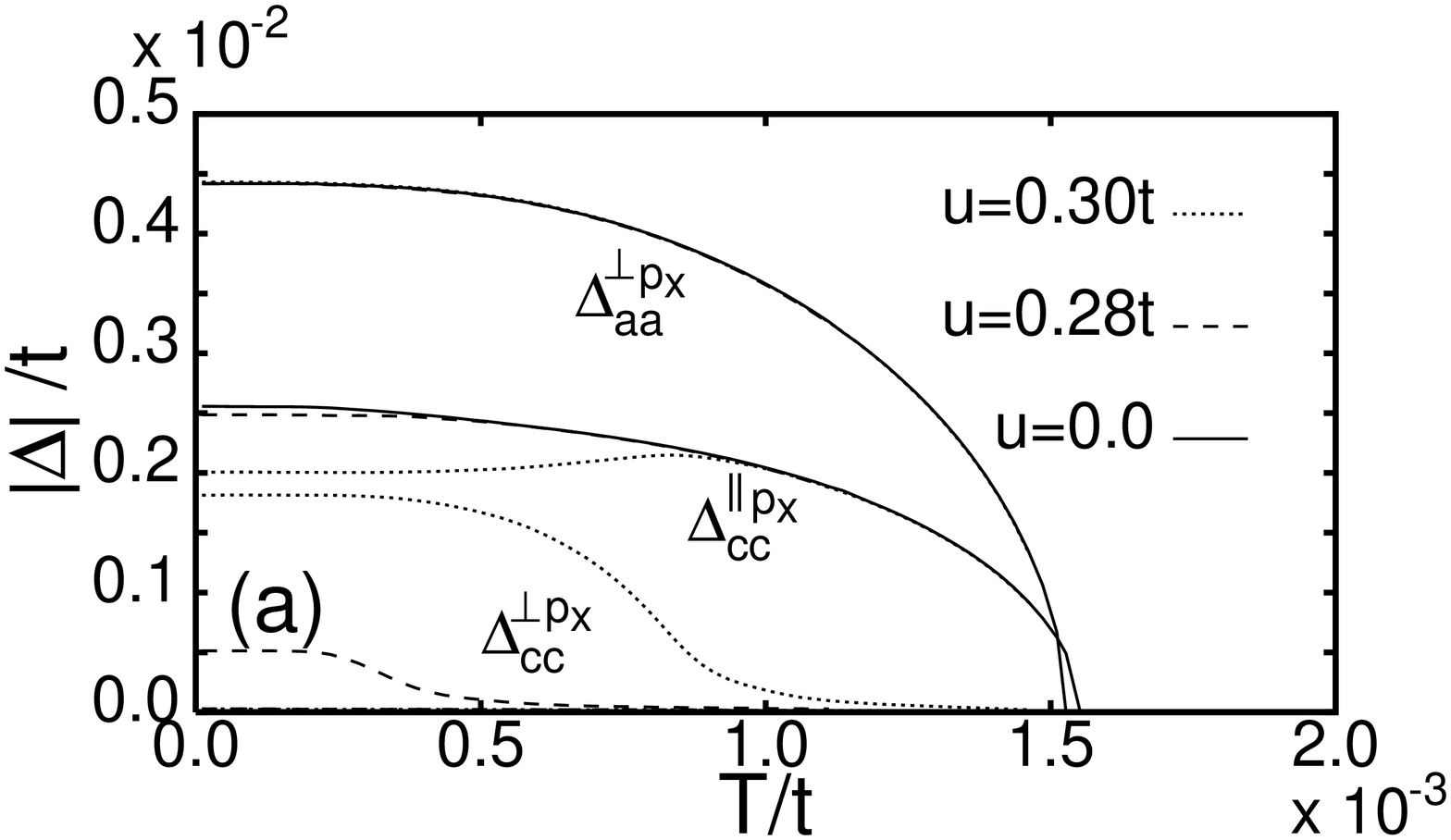,width=4.5cm,angle=-90}}
\vspace{0.5cm}

\centerline{\epsfig{file=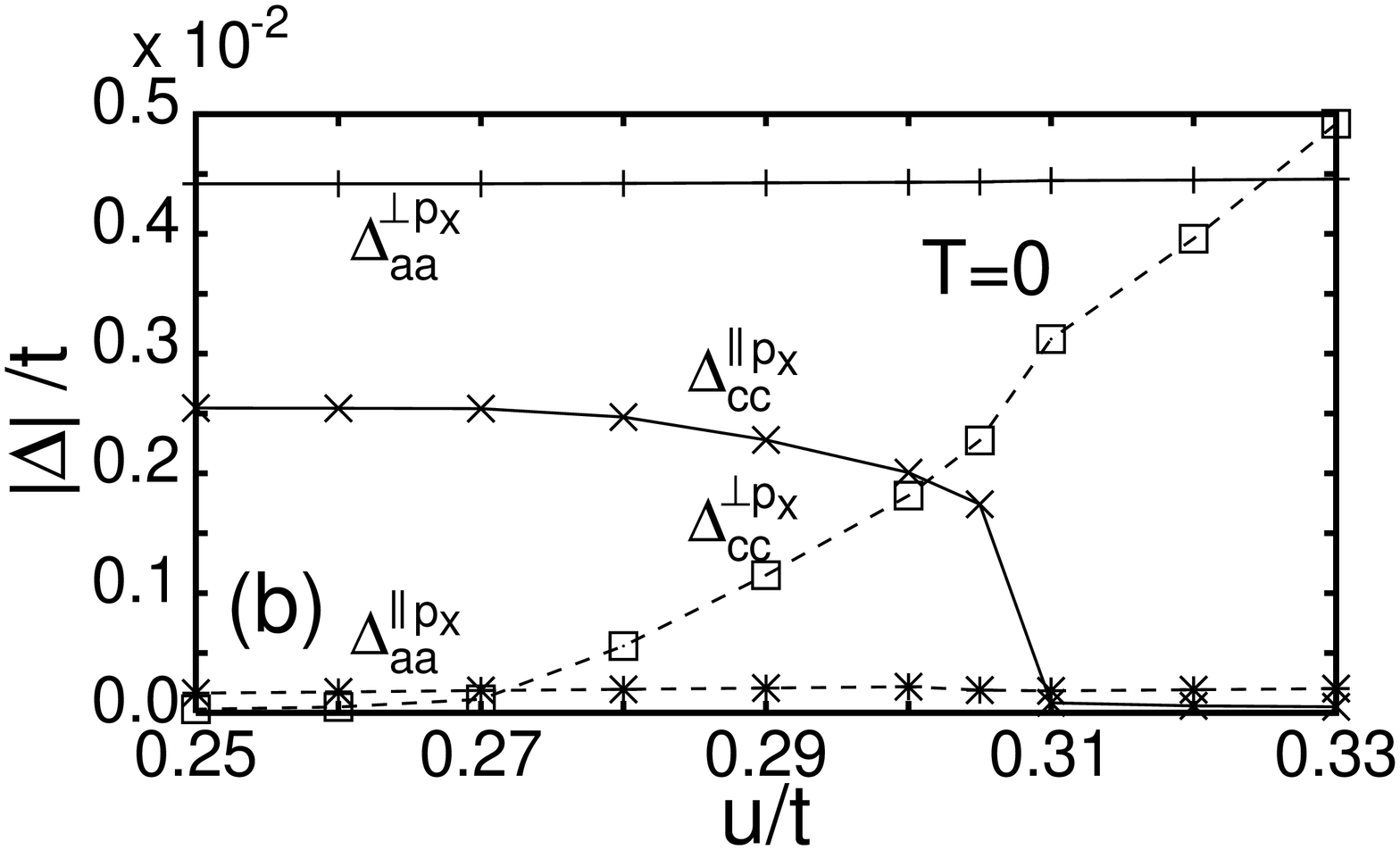,width=4.5cm,angle=-90}}
 \caption{\label{fig_three} (a) Temperature dependence of
  order parameters $|\Delta^{\perp p_x}_{aa}|$,
$|\Delta^{\parallel p_x}_{cc}|$ and $| \Delta^{\perp p_x}_{cc}|$
for a number of $u$ values ($u'=u$).  (b) Order parameters
$|\Delta^{\parallel p_x}_{cc}|$ and $|\Delta^{\perp p_x}_{aa}|$,
$|\Delta^{\perp p_x}_{cc}|$ at zero temperature versus the
interaction parameter $u(=u')$.}
 \end{figure}

\begin{figure}[bh]
\centerline{\epsfig{file=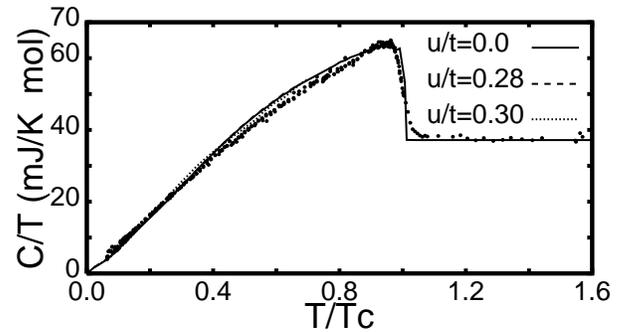,width=4.5cm,angle=-90}}
\caption{\label{fig_four} Calculated specific heat for a three
parameters $u(=u')$ ($u/t=0.0,~0.28$ and 0.30 corresponding to full
dashed and dotted lines, respectively) compared to the experimental
data (points) of NishiZaki {\it et al.}.[8]}
\end{figure}

It is important to ask how these results depend on the details of
the assumptions made in the model. In order to test the stability
of our results to variations in the model we  therefore introduced
some
 additional subdominant interaction
parameters.  For our initial exploration of the issues involved we
have generalized Eqs. (\ref{eq21}) and (\ref{eq22}) as follows:
\begin{eqnarray}
U_{\parallel mm'}= \left( \begin{array}{ccc} u & u & u \\
u & u & u \\
u & u & U_{\parallel}
   \end{array} \right) \label{new_u_set} \\
U_{\perp mm'}= \left( \begin{array}{ccc} U_{\perp} & U_{\perp} & u' \\
U_{\perp} & U_{\perp} & u' \\
u' & u' & u'
   \end{array} \right), \nonumber
\end{eqnarray}
  Reassuringly,
with these modified parameters we obtained a temperature
dependence of the gap parameters which are qualitatively similar
to those for the original parameters. 
It is interesting to note that for fixed values of
$U_{\perp}$ and $U_{\parallel}$ the changes of $u$ and $u'$
hardly change the superconducting transition
temperature. 
We have systematically studied the effect of 
additional interactions, especially so on the line $u=u'$, and found    
small differences compared to the $u=0$ solution even for  
 $u$  as large as $0.28t$ .
The  differences are  mainly connected with the appearance of
 out of plane
components of $\Delta^\perp_{cc}$ generated by the new
interactions as is evident from Fig.(\ref{fig_three}). 
For larger values of $u$ the difference becomes more significant
(Fig.~\ref{fig_three}a). Note, however, that only low temperature
dependence of the pairing amplitudes is affected. In
Fig.~\ref{fig_three}b we show the variation of a few characteristic
$|\Delta_{mm}|$ against $u$  at zero temperature. Clearly, for $u
> 0.3t$ there is a qualitative change of our solution leading to
dominant out of plane pairing components in all orbitals. Large
$u$  also affects the critical temperature $T_c$. Interestingly,
for finite $u$ we also observe increasing values
 of in-plane pairing
amplitudes in the $a$ and $b$ channels: $\Delta_{m,m'}^{\perp
p_x}$ and $\Delta_{m,m'}^{\perp p_y}$ 
for $m,m'=a,b$. Reassuringly, the corresponding specific heat
(Fig.~\ref{fig_four}) is essentially unchanged and remains 
in equally good
agreement with the experiments. Therefore we conclude that the
solution we have found is not very specific to the precise details
of the model parameters which we assumed, but is a generic
solution valid for at least some range of the possible interaction
parameters of the form depicted in Eq. \ref{new_u_set}.

\begin{figure}[thb]
\vspace{1.0cm}
\centerline{\epsfig{file=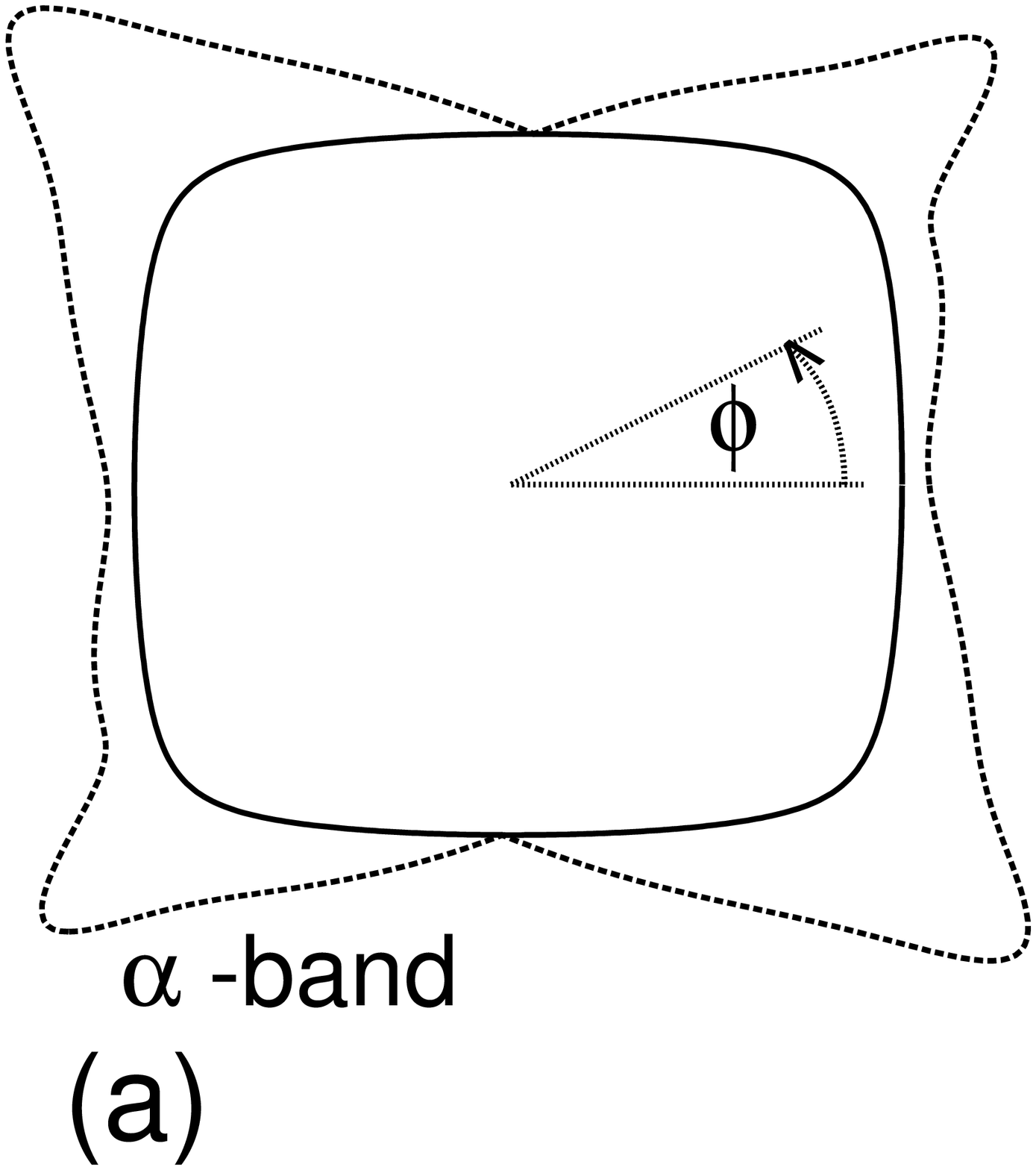,width=3.5cm,angle=-90}
\hspace{-0.8cm}
\epsfig{file=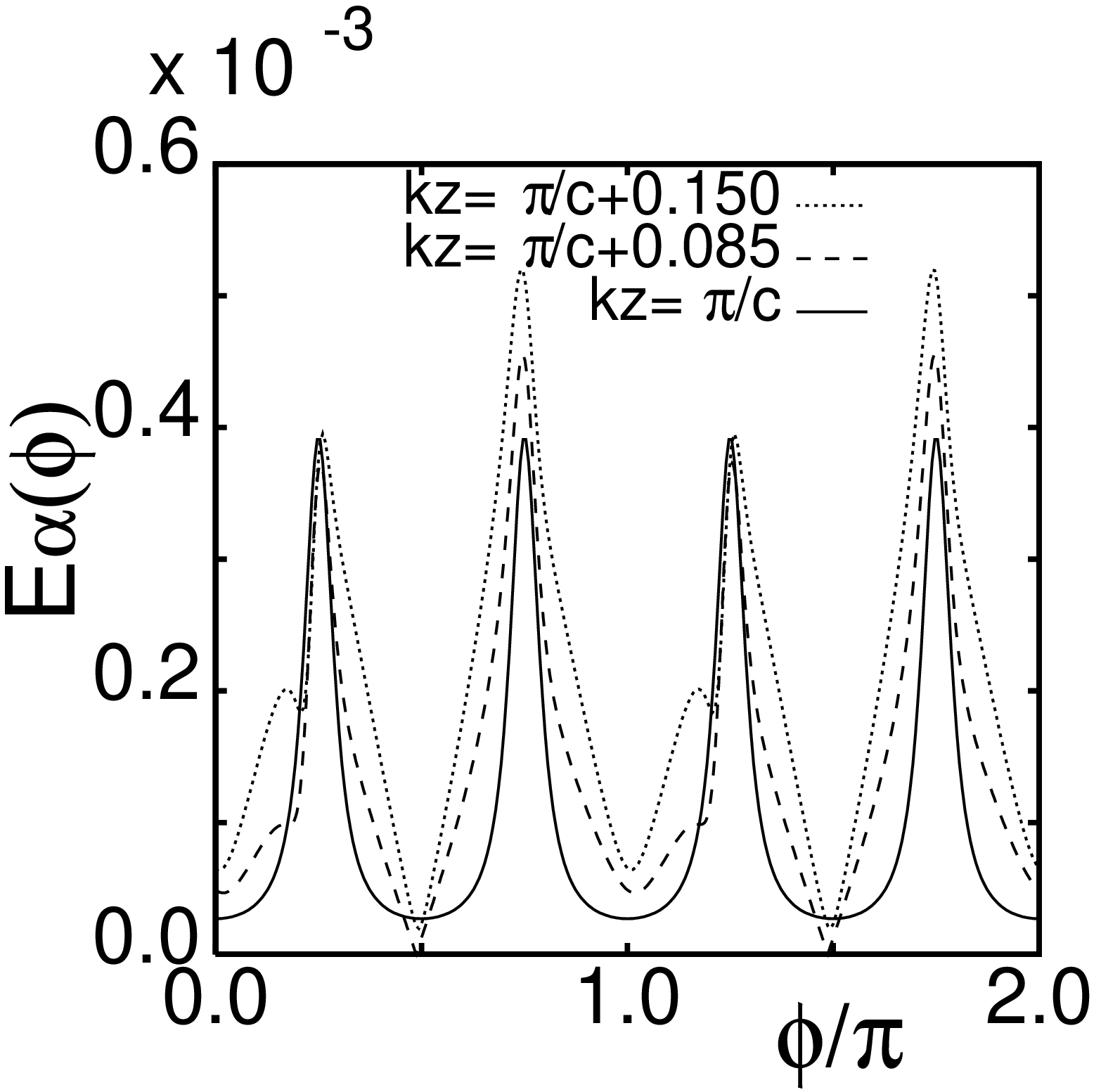,width=4.5cm,angle=-90}}
\vspace{-0.5cm}

\centerline{\epsfig{file=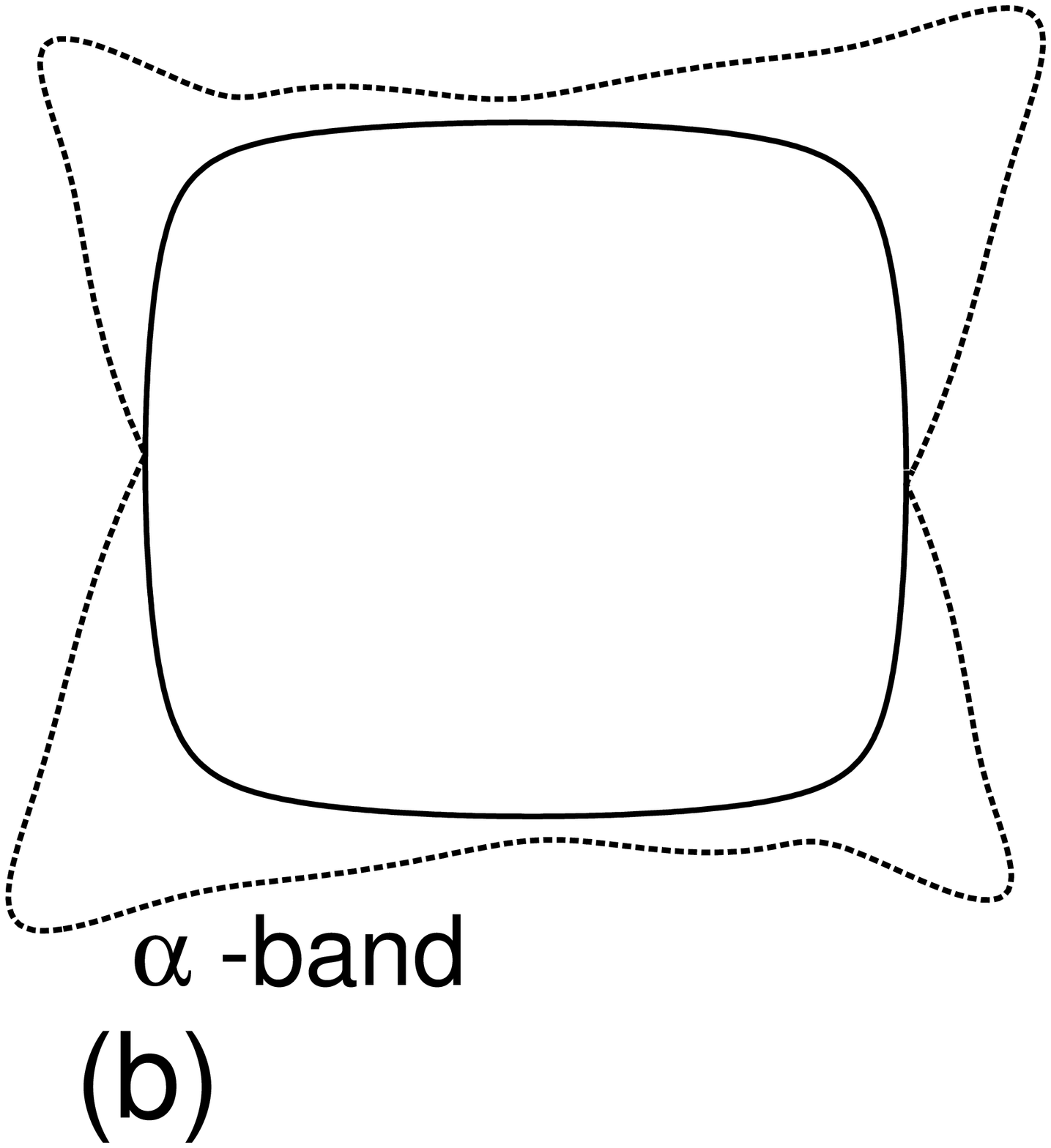,width=3.5cm,angle=-90}
\hspace{-0.8cm}
\epsfig{file=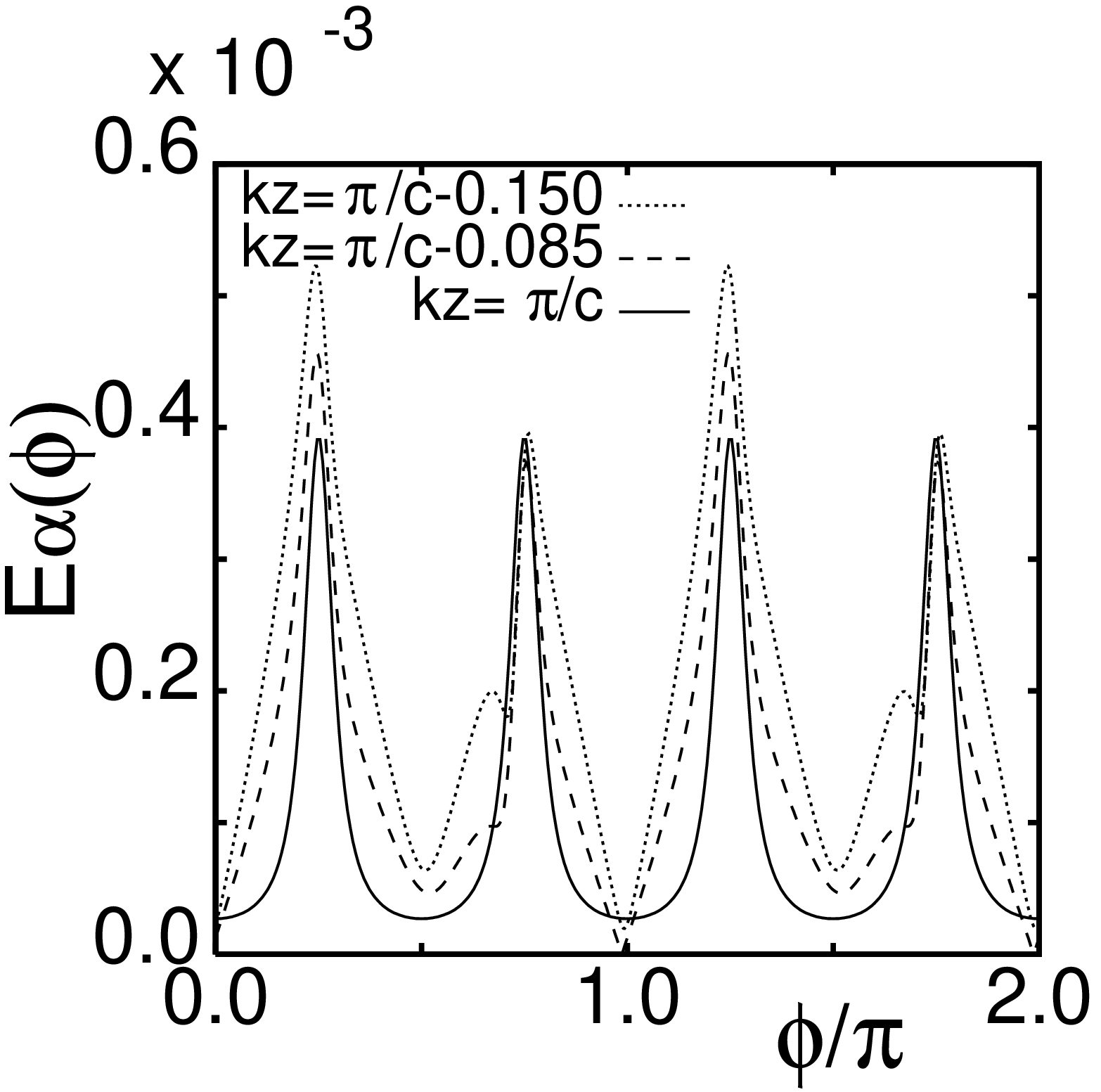,width=4.5cm,angle=-90}}
\vspace{-0.5cm}

\centerline{\epsfig{file=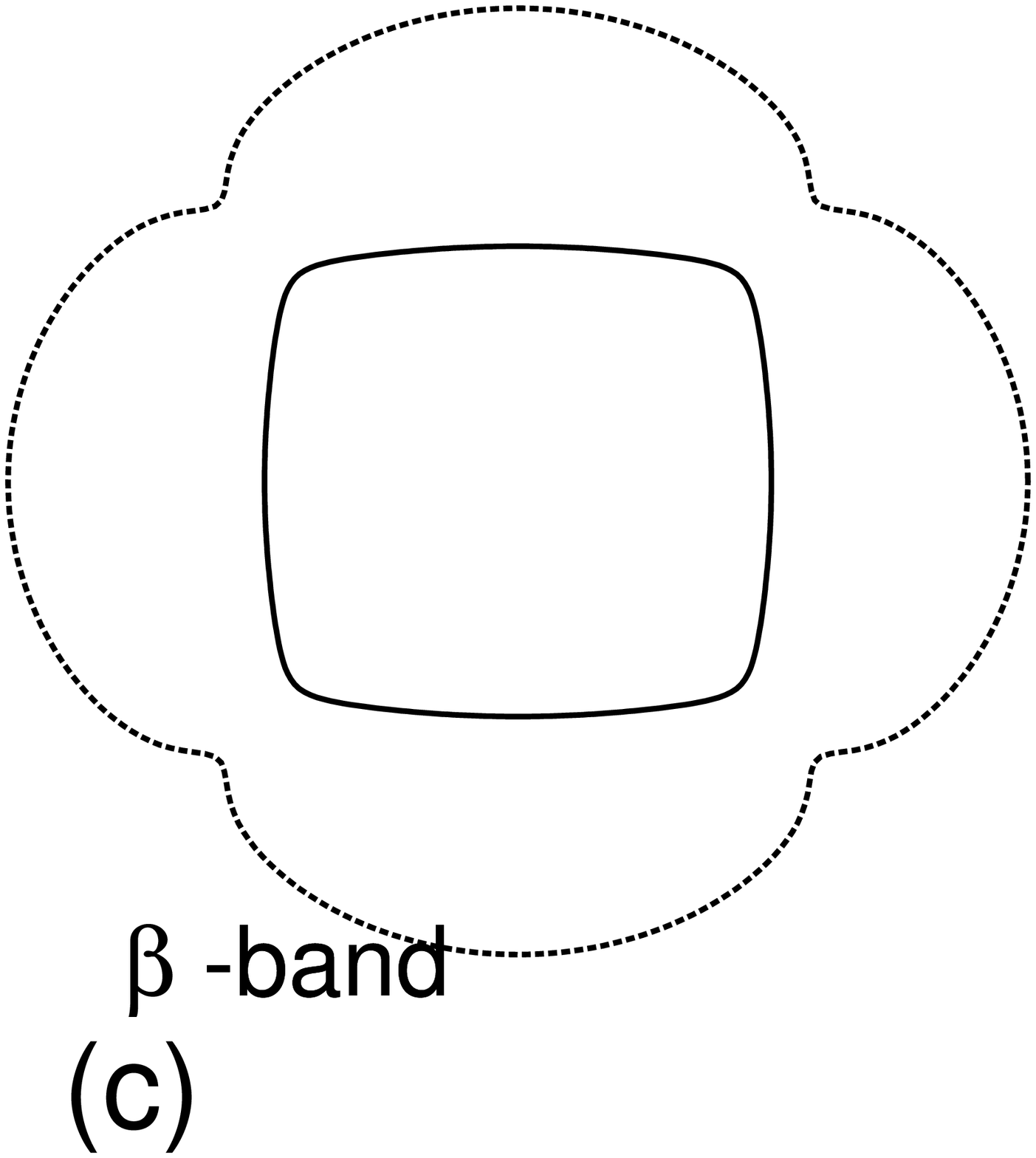,width=3.5cm,angle=-90}
\hspace{-0.8cm}
\epsfig{file=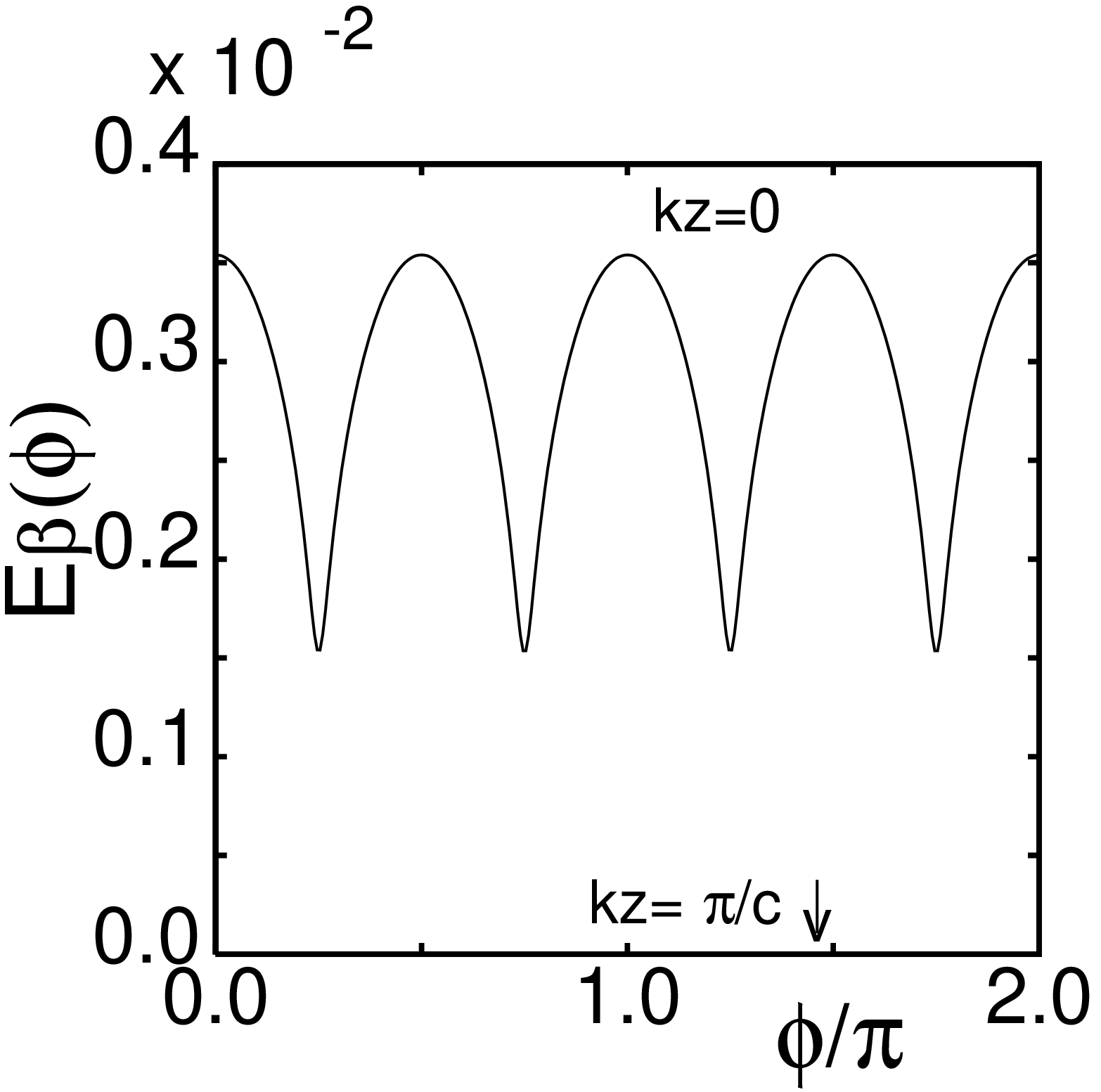,width=4.5cm,angle=-90}}
\vspace{-0.5cm}

\centerline{\epsfig{file=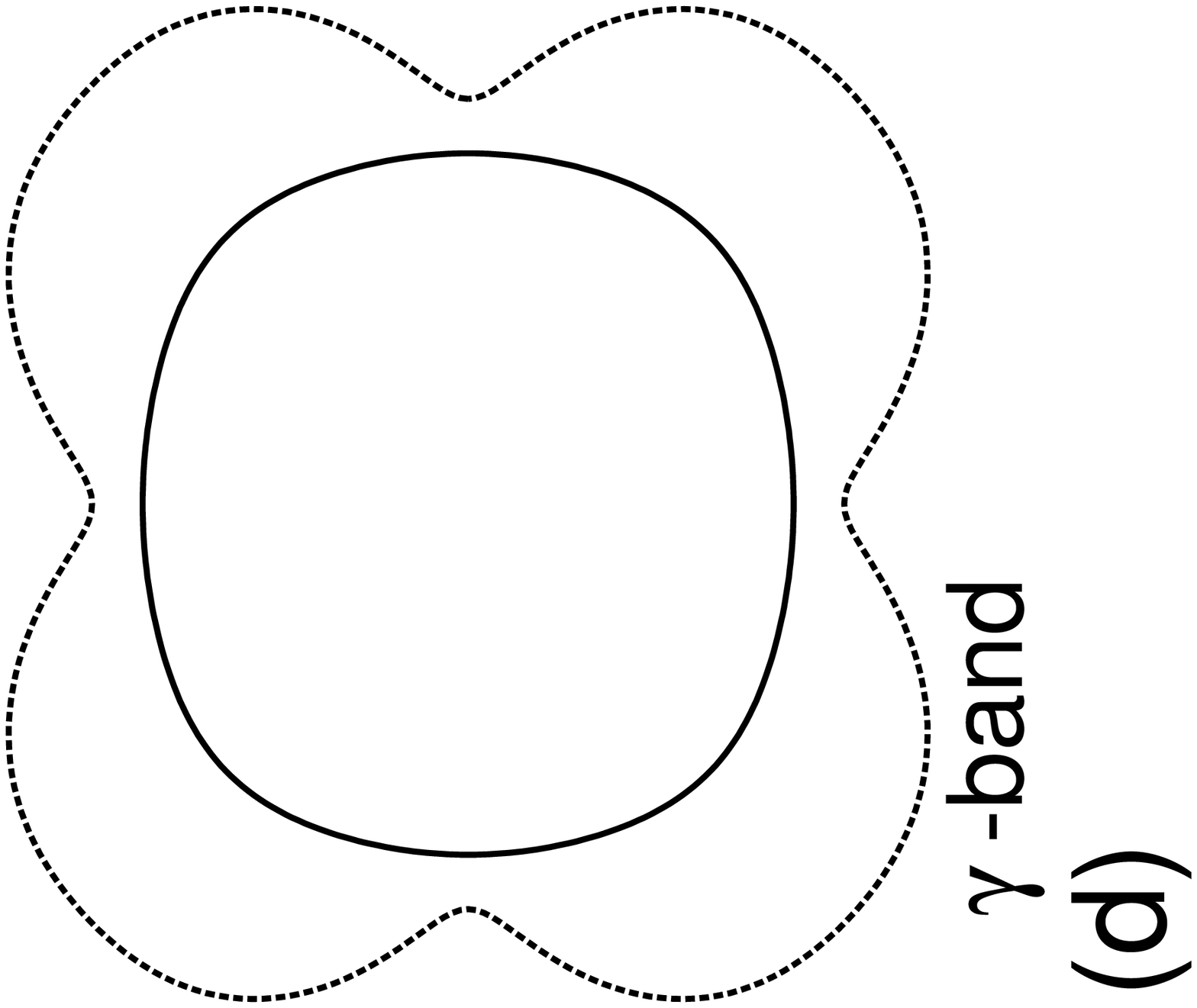,width=3.5cm,angle=-90}
\hspace{-0.8cm}
\epsfig{file=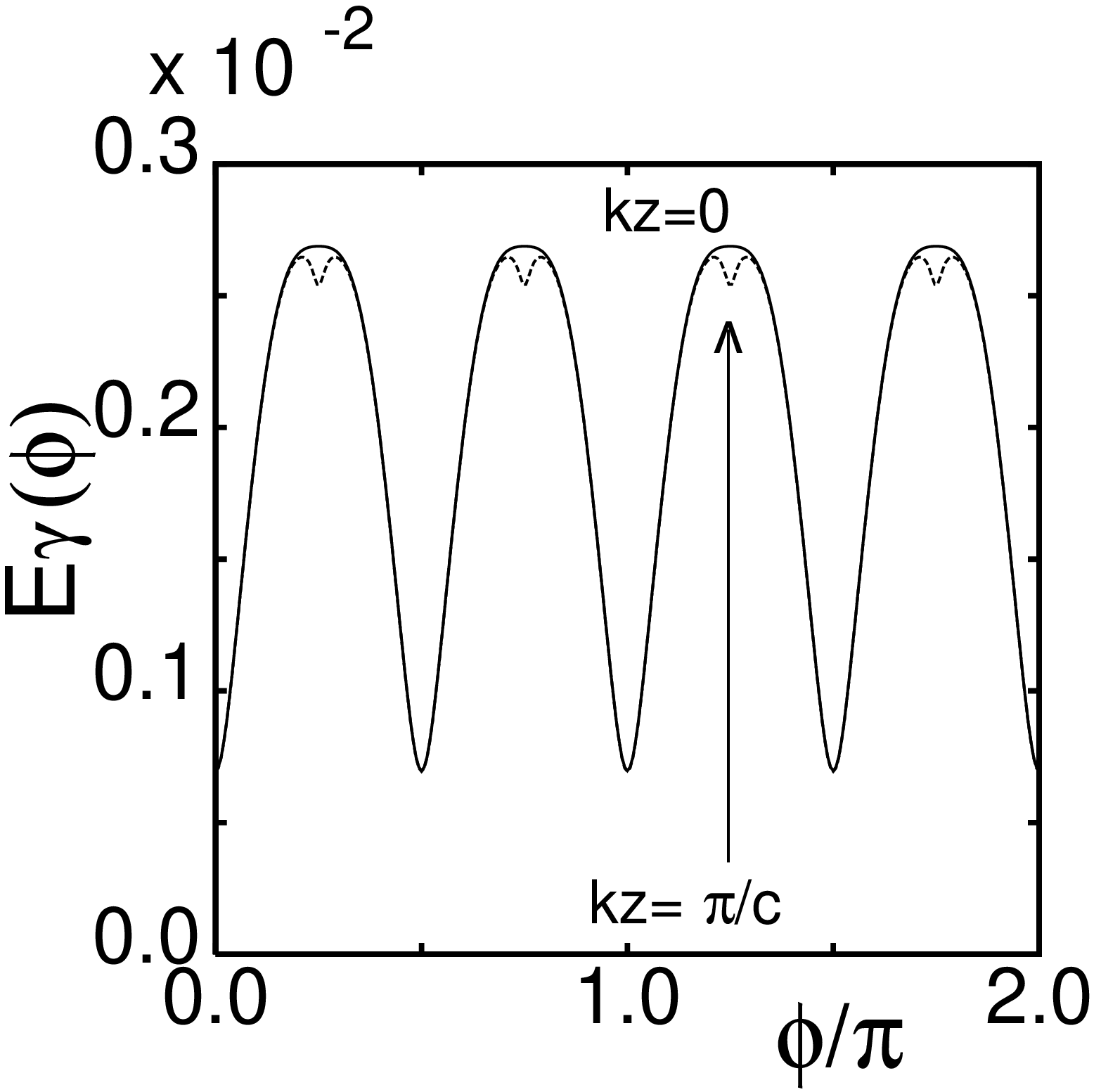,width=4.5cm,angle=-90}}

\caption{\label{fig_five}Lowest energy eigenvalues, $E^\nu({\bf
k})$ on the Fermi surface; $\alpha$  sheet in the plane
$k_z=\pi/c+0.085$ (a) and $k_z=\pi/c+0.085$ (b), $\beta$ (c) and $\gamma$ (d) sheets in the
plane $k_z=0$.
}
\end{figure}

The quasiparticle energy gap structure which we obtained is shown in
Fig.~\ref{fig_five}.
The gap is finite everywhere on the $\gamma$ sheet,
Fig.~\ref{fig_five}(d), although
it is very anisotropic, and becomes
small when the Fermi surface approaches near to the
van Hove points at $(\pi,0)$ and $(0,\pi)$.
In contrast, the $\alpha$ and $\beta$
Fermi surface sheets have gap zeros in the vicinity of the lines
$k_z=\pm \pi/c$. In the case of $\beta$ the gap is zero to numerical accuracy
on these nodal lines.  While in the case of $\alpha$ the gap is very small
on these lines, but not exactly zero. In fact there are eight point nodes on the
$\alpha$ sheet, as can be seen in Fig.~\ref{fig_five}(a,b). Two point nodes lie just
above
 the $k_z= \pi/c$ line at $k_z \approx \pi/c+0.085$ at
two different angles. Another pair lie just below, at $k_z \approx
\pi/c -0.085$ at an angle rotated by $\phi=\pi/2$. The remaining
four are located in similar positions near the line $k_z=-\pi/c$.
This interesting nodal structure arises from the fact that the
$\alpha$ Fermi surface cylinder is centered at $X$ in the
Brillouin zone not at $\Gamma$ (Fig.~\ref{fig_two}), and therefore
it has two-fold symmetry not four fold like $\beta$ and $\gamma$.
Notice also that the excitation gap on the $\alpha$ sheet is
non-zero even when $\Delta_{aa}=\Delta_{ab}=\Delta_{bb}=0$,
because it is hybridized to the $c$ orbital and $\Delta_{cc}\neq
0$.

\begin{figure}[thb]
\vspace{0.0cm}
\centerline{\epsfig{file=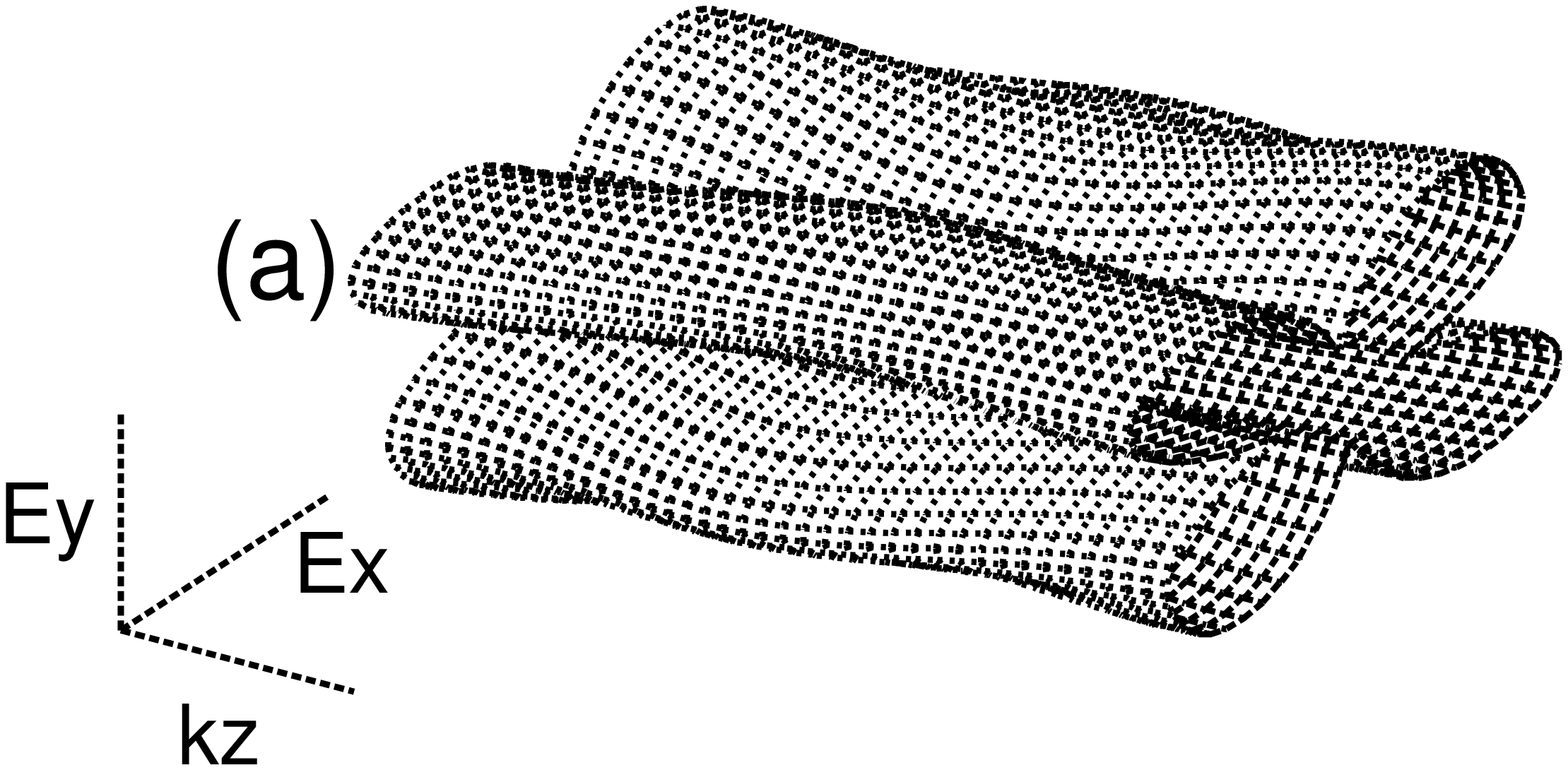,width=5.5cm,angle=-90}}
\vspace{-3.0cm}
\centerline{\epsfig{file=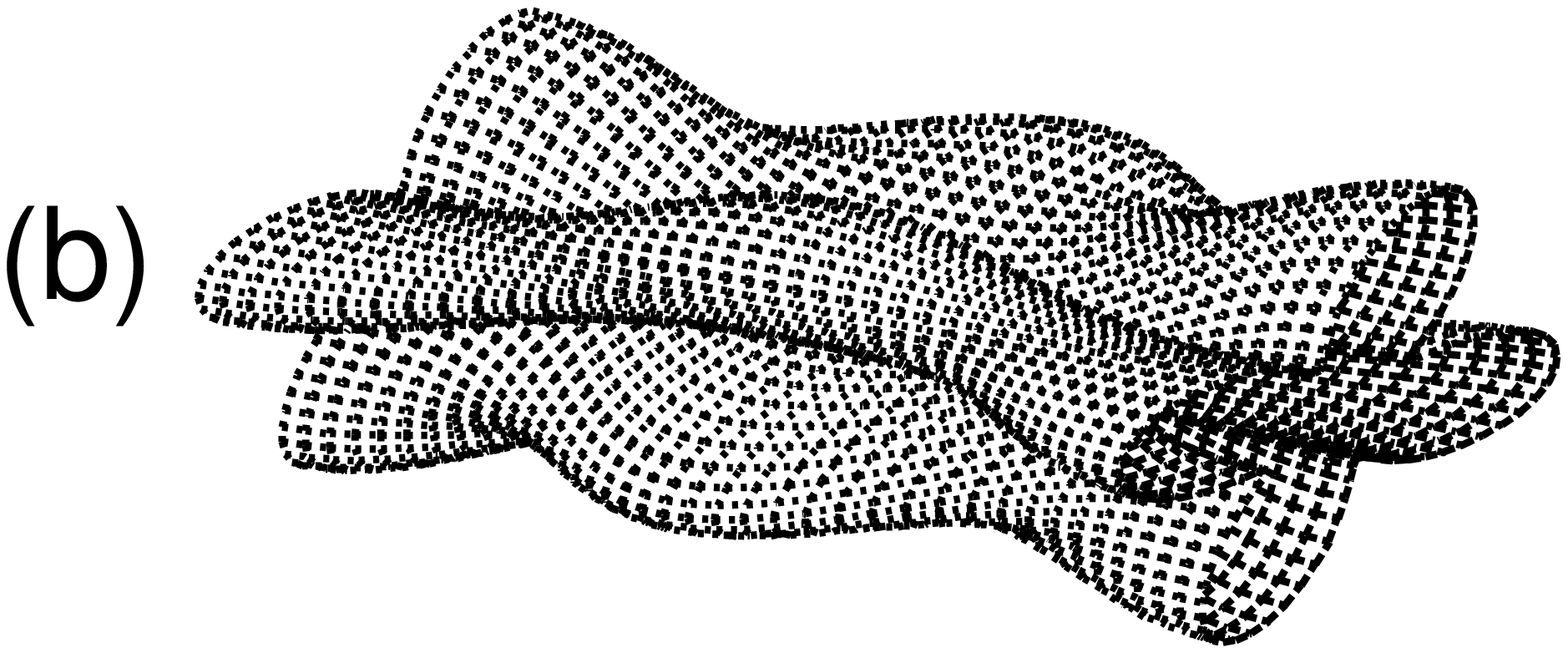,width=5.5cm,angle=-90}}
\vspace{-3.0cm}
\centerline{\epsfig{file=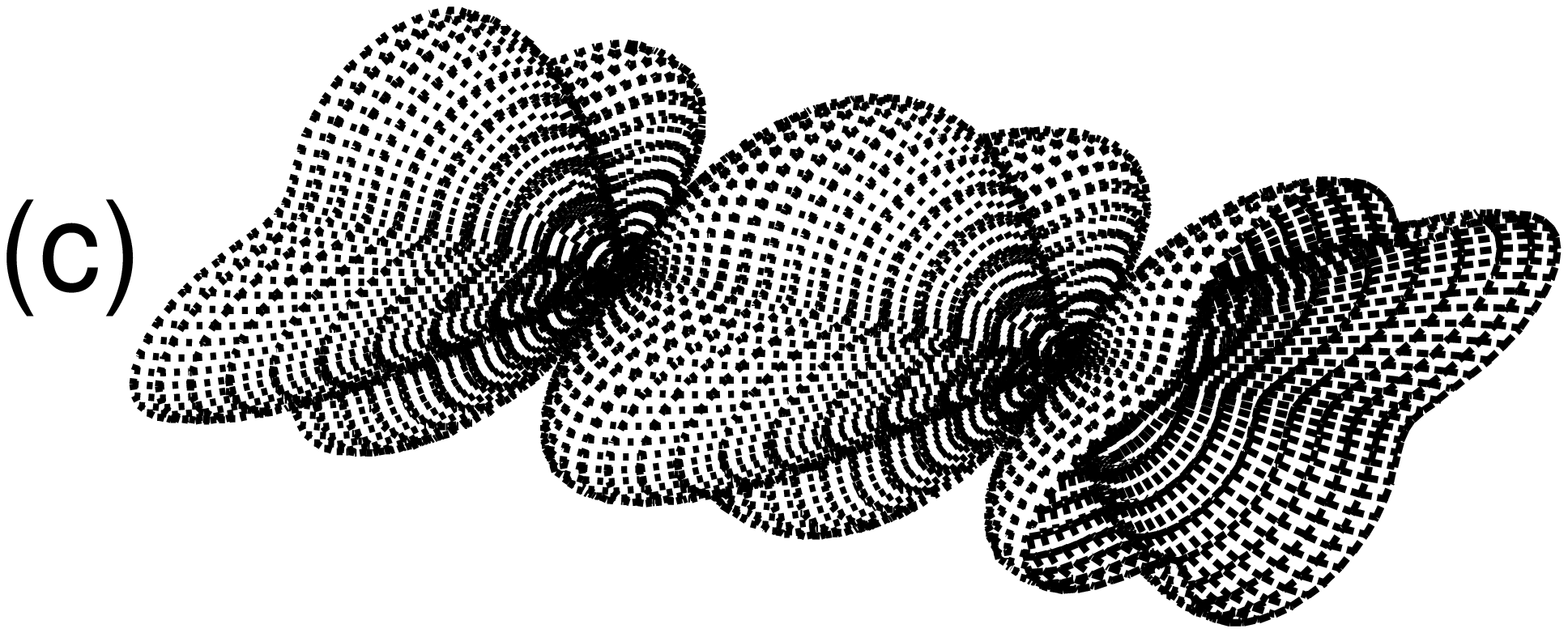,width=5.5cm,angle=-90}}
\vspace{-2.0cm} \caption{\label{fig_six} Minimum energy
quasiparticle eigenvalues on the $\gamma$ Fermi surface sheet,
$E({\bf k_F})$, plotted in cylindrical polar coordinates as
functions of $k_z$ and $a-b$ plane polar angle, $\theta$.
Parameter values are $u/t=0.28$ (a), 0.30 (b), 0.32 (c),
respectively. One can see that for $u \leq 0.3$ the $\gamma$ sheet
gap is nodeless, while for $u > 0.3$ line nodes appear at $k_z=\pm
\pi/c$.}
\end{figure}

Note that this nodal structure of the gap is unchanged by the
presence of the small  subdominant interaction parameter $u$, in
Eq.(\ref{new_u_set}). However, upon increasing the value of the $u$
parameter eventually the results change qualitatively, leading to
appearance of additional line nodes in  $\gamma$ (Fig.
(\ref{fig_six})) for $u=0.32t$. In this case the $\gamma$ band
gap also develops a line node, similar to the behavior of the
$\beta$ band.


\section{Effects of Disorder}

As we noted it earlier,  to obtain agreement with experiment we had
to eliminate the $f$-wave component $\Delta^{\perp f}_{mm'}(T)$
and we suggested that this can be done by an appeal to the effects
of a small amount of disorder. We shall now substantiate this
contention by explicit calculations.

 In case of non-magnetic disorder our Hamiltonian can be
written
\begin{eqnarray}
  \hat{H}& =& \sum_{ijmm',\sigma}
\left( (\varepsilon_m + \epsilon_i - \mu)\delta_{ij}\delta_{mm'}
 - t_{mm'}(ij) \right) \hat{c}^+_{im\sigma}\hat{c}_{jm'\sigma} \nonumber \\
&& - \frac{1}{2} \sum_{ijmm'\sigma\sigma'} U_{mm'}^{\sigma\sigma'}(ij)
 \hat{n}_{im\sigma}\hat{n}_{jm'\sigma'} \label{hubbard_d}
\end{eqnarray}
where $\epsilon_i$ is a random site energy. For a given
configuration of  $\epsilon_i$ one can, in principle,  perform
calculations (Eq. \ref{deltas}-\ref{eq18}) and then average over many
configurations. More readily, for highly disordered systems it is
possible to apply mean field theory of disorder by making use of
the Coherent Potential Approximation CPA
\cite{mar99,lit00,lit01,lit02c}.

\begin{figure}
\centerline{\epsfig{file=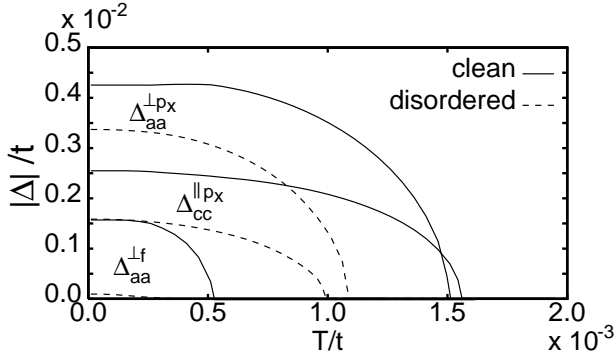,width=4.5cm,angle=-90}}
 \caption{\label{fig_seven}
  Order parameters $|\Delta^{\perp p_x}_{aa}|$,
$|\Delta^{\parallel p_x}_{cc}|$, $|\Delta^{\perp f}_{aa}|$ for the clean
and disordered systems ($\tau^{-1}
=0.005$)
 as functions of temperature.}
 \end{figure}

Here however, as superconducting Sr$_{2}$RuO$_{4}$ samples were
found to be relatively clean, we can limit our analysis to weak
disorder and non-resonant impurity scattering. Then knowing the
scattering rate $\tau^{-1}$ we can apply the Born approximation
\cite{agt99} 
 in
calculating the self-energy of the disorder averaged Green
function. Following this Abrikosov-Gorkov approach,  we assume
that impurity scattering will create a finite imaginary self
energy of the order
\begin{equation}
\Sigma({\rm i} \omega)= {\rm i} {\tau}^{-1} {\rm sgn} (\omega).
\end{equation}
Thus our equation of self-consistency in the configurationally
averaged pair potential can be written in terms of Matsubara
frequencies $\omega_n=(\pi/\beta)(2n+1)$ as follows
\begin{eqnarray}
 \Delta^{\sigma\sigma'}_{mm'}(ij) &=& U_{mm'}^{\sigma\sigma'}(ij)
\sum_{n=-\infty}^{\infty} {\rm e}^{{\rm i} \omega_n\delta} \label{eq26}\\
&\times&
\frac{1}{\beta} \sum_{\nu} \frac{ u^\nu_{im\sigma} v^{\nu *}_{jm'\sigma'} }{{\rm i} (\omega_n+ \tau^{-1}\omega_n/|\omega_n|) - E^{\nu}}, \nonumber
\end{eqnarray}
where $\delta$ denotes a positive infinitesimal. Exchanging the
summations over $\nu$ and $n$ indices  Eq. \ref{eq26} can be written,
after decoupling the summation over negative and positive
Matsubara frequencies,  as
\begin{eqnarray}
\Delta^{\sigma\sigma'}_{mm'}(ij) &=& -U_{mm'}^{\sigma\sigma'}(ij)
\frac{1}{\beta} \sum_{\nu} u^\nu_{im\sigma} v^{\nu *}_{jm'\sigma'} \\
&\times&
\sum_{n=0}^{\infty}
 \frac{ E^{\nu} }{ (\omega_n +\tau^{-1})^2 + (E^{\nu})^2 },  \nonumber
\label{eq27}
\end{eqnarray}

Conveniently the sum on the right hand site of Eq. \ref{eq29} can be
evaluated \cite{han75}
and it leads
to the final formula:
\begin{eqnarray}
\Delta^{\sigma\sigma'}_{mm'}(ij) &=& -U_{mm'}^{\sigma\sigma'}(ij)
\sum_{\nu} u^\nu_{im\sigma} v^{\nu *}_{jm'\sigma'} \\
&\times&
 \frac{1}{2 \pi} {\rm Im} \Psi \left( \frac{1}{2} + \frac{\beta}{2 \pi \tau} + \frac{{\rm i} E^{\nu} \beta}{2 \pi} \right). \nonumber
\label{eq28}
\end{eqnarray}

Note that in the limit of a clean system $\tau^{-1} \rightarrow 0$
\begin{equation}
 \frac{1}{2 \pi} {\rm Im} \Psi \left( \frac{1}{2} + \frac{\beta}{2 \pi \tau} + \frac{{\rm i} E^{\nu} \beta}{2 \pi} \right)
\rightarrow -(1 - 2 f(E^{\nu}))
\label{eq29}
\end{equation} and Eq. \ref{eq28}
 coincides with that of the clean system  (Eqs.
\ref{bogoliubov}-\ref{deltas}).

\begin{figure}
\centerline{\epsfig{file=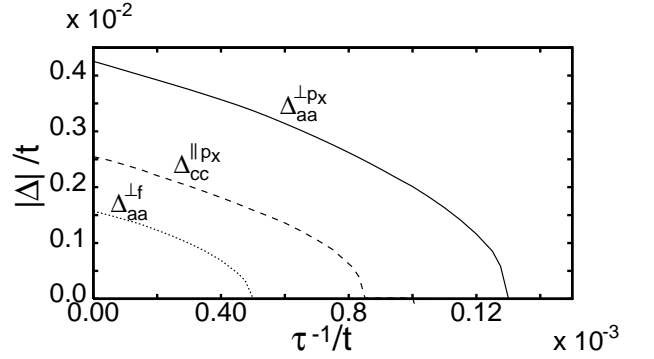,width=4.5cm,angle=-90}}
 \caption{\label{fig_eight}
  Zero temperature order parameters $|\Delta^{\perp p_x}_{aa}|$,
$|\Delta^{\parallel p_x}_{cc}|$, $|\Delta^{\perp f}_{aa}|$  as a function of
impurity scattering rate $\tau^{-1}$.}
 \end{figure}

Thus, in a  weak disorder limit we have again solved the
Bogoliubov-de Gennes equations including a small $\tau^{-1}$. The
results of our calculations are shown in Fig. \ref{fig_seven} where
we have plotted the order parameters $|\Delta^{\perp p_x}_{aa}|$,
$|\Delta^{\parallel p_x}_{cc}|$, $|\Delta^{\perp f}_{aa}|$ versus
temperature. Evidently in the disordered case the small $f$-wave
amplitude ($|\Delta^{\perp f}_{aa}|$) is reduced to zero much more
rapidly than the larger p-wave ones ( $|\Delta^{\perp p_x}_{aa}|$,
 $|\Delta^{\parallel p_x}_{cc}|$).
Furthermore, in Fig. \ref{fig_eight} we show disorder dependence
of these three paring amplitudes at zero temperature. Clearly each
order parameter is reduced to zero in a typical Abrikosov-Gorkov
like manner, becoming zero approximately when the pair-breaking
parameter $ \tau|\Delta^{clean}(0)|/\pi \approx 1 $.

From Figs. (\ref{fig_seven}) and (\ref{fig_eight}) it is clear that,
for moderate scattering rates, there is a region where the f-wave
gap components are reduced to zero but the larger p-wave
components are more or less unaffected. Thus we conclude that the
simultaneous neglect of $\Delta^{\perp p_z}_{mm'}$ and
$\Delta^{\perp f}_{mm'}$ and $\tau^{-1}$ is
justified\cite{annett01}.  It would be an interesting experimental
confirmation of this model, if ultra-clean samples were found to
have a second phase transition at a much lower temperature than
$T_c \sim 1.5{\rm K}$.

\section{Bond-proximity effects}

To get a single superconducting transition temperatures within
our interlayer coupling model we are forced to fine tune 
two interaction parameters. However, it has been 
proposed\cite{zhitomirsky01}
that a single transition can be obtained in an multiband model 
by allowing for a symmetry mixing interaction  of the type
$U({\bf k},{\bf q})=g'f({\bf k})g({\bf q})$, where 
$f({\bf k})$ and $ g({\bf q})$
are order parameter symmetry functions for respective bands. 

It is the aim of the present section
to check to which extend similar approach may be used 
in our bond model. 
We start with short discussion of the source and 
magnitude of symmetry mixing interaction. 
The description we have used is a real space, 
two point near neighbour interaction
such as naturally arises in any multi-orbital, extended, negative $U$
Hubbard model, Eq.~\ref{hubbard}. To be quite clear about this
matter we recall that a generic pair-wise interaction like $U({\bf
r},{\bf r}')$, when expressed in the language of a tight-binding
model Hamiltonian will, in general, give rise to four point
interaction parameters $U_{ij,kl}$. The original Hubbard
Hamiltonian makes use of the one point parameters
$U_{i}^{(1)}=U_{ii,ii}$ whilst the extended Hubbard model  is
based on two point parameters $U_{i,j}^{(2)}=U_{ij,ij}$. Evidently
our `bond' model is a negative U-version of the
latter\cite{micnas}. 
The  symmetry mixing interactions\cite{zhitomirsky01} 
arise from 3-site interactions $U_{i,j,l}^{(3)}$.
The physics of this is often referred to as assisted
hopping\cite{zawadowski}. If one assumes, as is normally the case
in an isotropic substance, that $
|U^{(1)}|>|U^{(2)}|>|U^{(3)}|>|U^{(4)}|$ then the `bonds'
represent stronger coupling than  assisted hopping  and should be
the preferred coupling mechanism. However, for the tetragonal
arrangement of Ru atoms in Sr$_{2}$RuO$_4$ this is no more than a
suggestion at present. 

In the presence of a three point interaction
$U_{i,j,l}^{(3)}=U_{ij,il}=U_{I}$, for all nearest neighbours
$ijl$ such that $i$ and $j$ are in one Ru plane while $l$ is on a
neigbouring one (Fig. \ref{fig_one}), the gap equation
(Eqs.~\ref{bogoliubov}-\ref{eq18}) can be rewritten in k-space
as,\cite{lit02}

\begin{eqnarray}
\Delta^{\sigma\sigma'}_{mm'}({\bf k}) &=&
\frac{1}{N}\sum_{\bf q} U_{mm'}^{\sigma\sigma'}({\bf k}- {\bf q})
\chi_{mm'}^{\sigma\sigma'}({\bf q}) \nonumber \\
&+&  \frac{1}{N} \sum_{{\bf q}, oo'}
U_{mm',oo'}^{\sigma\sigma'}({\bf q},{\bf k} -{\bf q})
\chi_{oo'}^{\sigma\sigma'}({\bf q})\,. \label{eq30}
\end{eqnarray}
where, as before,
\begin{equation}
\chi_{mm'}^{\sigma\sigma'}({\bf k}) =
u^{\nu}_{{\bf k}m\sigma}v^{\nu *}_{{\bf k}m'\sigma'}
(1 - 2f(E^{\nu})\,.
\label{eq31}
\end{equation}

\begin{figure}
\centerline{\epsfig{file=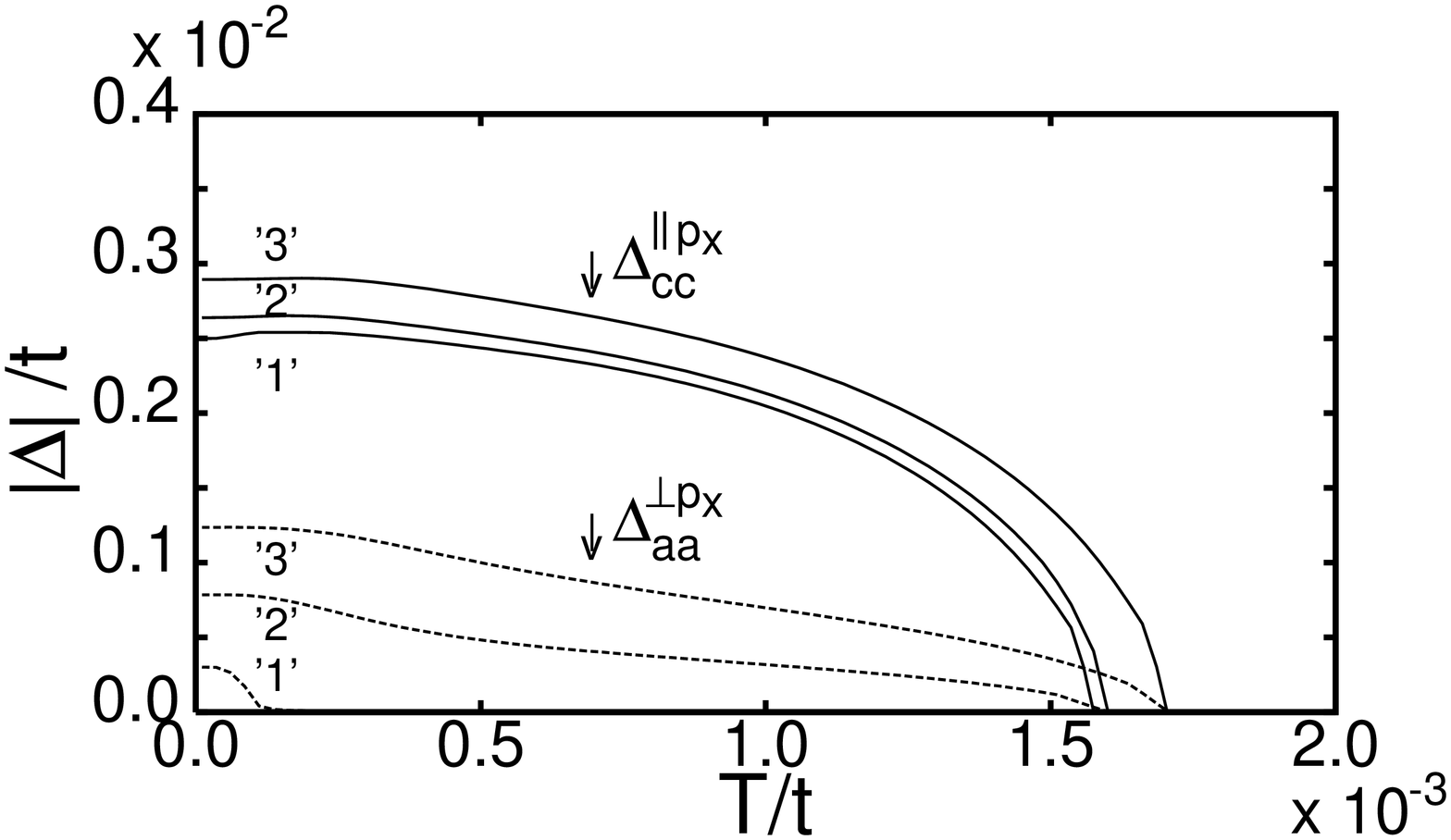,width=4.5cm,angle=-90}}
 \caption{\label{fig_nine}
  Order parameters $|\Delta^{\perp p_x}_{aa}|$,
$|\Delta^{\parallel p_x}_{cc}|$, for $U_{\perp}=0.400t$,
$U_{\parallel}=0.494t$ and
the
proximity coupling
 as functions of temperature.
The three curves '1', '2' and '3' correspond to the values
0.0, 0.005, 0.010 of
three point coupling constant.}
 \end{figure}

\begin{figure}[bh]
\centerline{ \epsfig{file=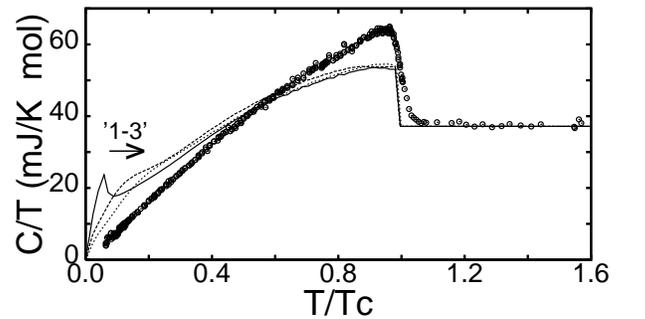,width=4.5cm,angle=-90}}
\caption{ \label{fig_ten} Specific heat and the effect three point
interaction, compared to the experimental data of NishiZaki {\it
et al.} [8]. The arrow indicates  increasing  values of the
proximity coupling ('1-2-3' as in Fig. \ref{fig_nine}). }
\end{figure}

In a body centered tetragonal crystal (Fig.~\ref{fig_one}) the
various matrix elements of the general four point interaction
$U_{mm',oo'}$ responsible for p-wave paring can be written
(suppressing spin indices for clarity):
\begin{eqnarray}
&& U_{cc}({\bf k},{\bf q}) = 2 U_{\parallel} V({\bf k})  V({\bf
q})
\label{eq32} \\
&&  U_{mm'}({\bf k},{\bf q}) = 8 U_{\perp} \tilde V({\bf k})
\tilde V({\bf q}) ~~~{\rm for}~~ m,m'=a,b \nonumber \\
&&
U_{mm',cc}({\bf k},{\bf q}) = 8 U_{I} \tilde V({\bf k}) V({\bf
q}) ~~~{\rm for}~~ m',m=a,b \nonumber \\
&& U_{cc,mm'}({\bf k},{\bf q}) = 8 U_{I} V({\bf k})  \tilde V({\bf
q}) ~~~{\rm for}~~ m',m=a,b \nonumber
\end{eqnarray}
where $V({\bf k})$ and $\tilde V({\bf k})$ are respectively:
\begin{eqnarray}
&& V({\bf k})= \left( \sin{k_x} +  \sin{k_y}\right)
\label{eq33}
\\
&& \tilde V({\bf k})= \left(\sin{\frac{k_x}{2}}
\cos{\frac{k_y}{2}} +  \sin{\frac{k_y}{2}}
\cos{\frac{k_y}{2}} \right) \cos{\frac{k_zc}{2}} \, \nonumber.
\end{eqnarray}

Note that the three point interaction leads to an extra interlayer
coupling proportional to $U_I$. Interestingly, the general form of
the order parameter is the same as previously derived (Eqs.
(\ref{bogoliubov}-\ref{eq20})) despite the additional three point
coupling Eq. (\ref{eq33}) in the self-consistency relation Eqs.
(\ref{eq31}-\ref{eq32}).
 
It has to be noted that the presence of interaction $U_I$ strongly 
changes $T_c$. To get its correct value (1.5K) for the present model
we have taken  $U_{\perp}=32$meV $  and $ $U_{\perp}=40$meV and
repeated our calculations for various $U_I$ values. The results are shown
in Figs. (\ref{fig_nine}) and (\ref{fig_ten}).

 Fig. (\ref{fig_nine}) shows the results for the
amplitudes $\Delta^{\parallel p_x}_{cc}(T)$
and $\Delta^{\perp p_x}_{aa}(T)$ including the three-point
interaction. Note that for $U_I =0$ 
(curves labeled by (1) in the figure) 
the temperature where $\Delta^{\parallel
p_x}_{cc}(T)$ becomes non-zero is much
higher than that where $\Delta^{\perp p_x}_{aa}(T)$
becomes non-zero. It is evident  from the figure that for $U_I \neq 0$ the
parameters $\Delta^{\Gamma}_{mm'}(T)$ for $a$, $b$ and $c$ orbitals
vanish at the same temperature and two transitions merge 
into one. Thus, the
proximity coupling mechanism identified by ZR\cite{zhitomirsky01}
 in their band
description of the electron-electron interaction, also works in
our bond model.

Intriguingly, although the proximity coupling works in principle,
the above mechanism does not seem to be helpful in the context of
building phenomenological interactions suitable to describe 
experimental data. To illustrate
this point we reproduce, in Fig. (\ref{fig_ten}), the specific heat
corresponding to the set of $\Delta^{\parallel p_x}_{cc}(T)$ and
$\Delta^{\perp p_x}_{aa}(T)$ shown in Fig. (\ref{fig_nine}). Clearly, as
$U_I$ increases the second transition at low temperature becomes
enlarged and merges with the first transition at higher
temperature. However, the small values of $U_I$ shown in Figs. (\ref{fig_nine}, \ref{fig_ten}) 
are not sufficient to get the specific heat jump at $T_c$ right. Therefore $U_I$ must be large for 
the ZR scenario to fit the experiments. 
By contrast, as we have demonstrated earlier,
if $U_I = 0$ and the sizes of $U_\perp$ and $U_\parallel$ are
adjusted so that only one transition occurs both the low
temperature slope and the jump at $T_c$ agrees with experiments.
Thus although we have not investigated models featuring a
`proximity effect' induced by $U^{(3)}$ type of interactions
systematically we conclude that such interactions are not needed
to fit the available data.

\begin{figure}
\centerline{ \epsfig{file=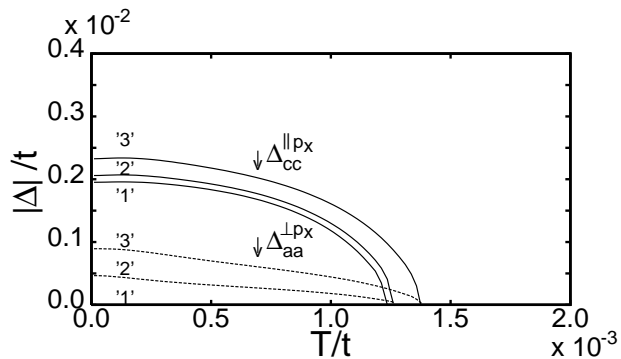,width=4.5cm,angle=-90}}
 \caption{\label{fig_eleven}
Order parameters $|\Delta^{\perp p_x}_{aa}|$, $|\Delta^{\parallel
p_x}_{cc}|$, for
$U_{\perp}=0.400t$,
$U_{\parallel}=0.494t$
 and
the proximity coupling
 as functions of temperature in disordered system $\tau^{-1}=0.005t$
The three curves '1', '2' and '3' correspond to the values
$U_I/t$: 0.0, 0.005, 0.010 of three point coupling constant. }
 \end{figure}

Finally,  it is also interesting to see what are the effects of
disorder on the `orbital proximity effect';  the results are shown
in Fig. \ref{fig_eleven}. We see that, disorder can eliminate the gap
on $\alpha$, $\beta$ sheets, while leaving it almost unchanged on $\gamma$.
This feature of the proximity effect scenario opens it up for experimental verification by 
measurement on samples with increasing disorder. Evidently the effect of disorder should be that 
the low temperature power laws disappear due to the destruction of superconductivity on the $\alpha$, 
$\beta$ sheets.


\bigskip
\noindent
{\bf VI. Conclusions}

\bigskip
We have introduced a methodology for building
semi-phenomenological, attractive electron-electron interactions
bond by bond for calculating superconducting properties under
circumstances when the physical mechanism of pairing is not known.
We deployed it to study $p$-wave pairing in Sr$_2$RuO$_4$. A bond
was described by an interaction constant
$U^{\sigma\sigma'}_{m,m'}(ij)$ which depends on the sites $i$ and
$j$, the orbitals $m$ and $m'$, and their spin orientation
$\sigma$ and $\sigma'$. We have solved the appropriate Bogoliubov
de Gennes equations for a number of scenarios defined by a small
set of interaction constants. We have found that the one for which
$U^{\uparrow\downarrow}_{cc}(ij) = U_\perp$ for $i$ and $j$ being
nearest neighbour Ruthenium atoms in the Ru-O planes and
$U^{\uparrow\downarrow}_{aa}(ij) = U^{\uparrow\downarrow}_{bb}(ij)
= U_\perp$ for $i$ and $j$ being nearest on neighbouring planes
explained most of the available experimental data. Namely, the
corresponding solution featured a gap function on the
$\gamma$-sheet of the form $\Delta_{cc}(\vec k) \sim \sin k_x +
{\rm i} \sin k_y$  and and a line of gap on the $\beta$ sheet. For
this scenario the requirement that there be only one transition at
$T_c \simeq 1.5$~K fixed both $U_\perp$ and $U_\parallel$ and
hence all further results could be regarded as quantitative
predictions of the model. Remarkably, the model gave a
satisfactory account of the data for the specific heat $C(T)$,
superfluid density $n_s(T)$ and the thermal conductivity
$\kappa(T)$. 

We have also investigated the stability of the model
to introduction of further interaction constants and disorder. We
found that the predictions of the model are robust to  
changes of new interactions, while disorder mainly affects
the f-wave solution. Thus we can conclude that 
the experimental data support
a simple model which describes, quantitatively, the $p$-wave
pairing observed in Sr$_2$Ru0$_4$ on the basis of two orbital
specific coupling constants: $U_\parallel = 40$~meV $U_\perp =
48$~meV. The central physical feature of the model is that
$U_\parallel$ corresponds to interaction between electrons in the
Ruthenium planes while $U_\perp$ describes an inter-plane
interaction of roughly equal strength.

In view of the above results, we would like to emphasize two
points. Firstly, we have proposed an alternative to the the
`intra-band proximity effect' model of Zhitomirsky and
Rice\cite{zhitomirsky01} for describing horizontal line nodes on
the $\alpha ,\beta $ sheets of the Fermi Surface in
superconducting Sr$_{2}$RuO$_{4}$. Our bond model differs from
theirs in the way the interlayer coupling is implemented. 
The extension of the model in the spirit of ZR has also been
studied by allowing for 3-site interactions in the Hamiltonian. 
Even though the resulting 'bond proximity model' features single
 superconducting transition 
temperature the original model with fine tuned two interactions 
gives better fit to experimental T dependence of the specific heat.

\section*{Acknowledgments}

This work has been partially supported by KBN grant No. 2P03B 106 18, the
Royal Society Joint Project and
the NATO Collaborative Linkage Grant 979446. We are grateful to Prof. Y. Maeno
for providing us with the experimental specific heat data
reproduced in Figs. 4 and 10.


\begin{thebibliography}{99}
\bibitem{maeno01} Y. Maeno, T.M. Rice and M. Sigrist,
 Physics Today {\bf 54}, 42 (2001).
\bibitem{mackenzie03} A.P. Mackenzie, Y. Maeno,  Rev. Mod. Phys., {\bf 75} 
657 (2003). 
\bibitem{leggett75} A. J. Leggett, Rev. Mod. Phys., {\bf 47} 331 (1975).
\bibitem{mackenzie00} A. Mackenzie and Y. Maeno, Physica {\bf B280}, 148
 (2000).
\bibitem{bergemann00} C. Bergemann, S. R. Julian, A.P. Mackenzie, S.
NishiZaki and Y. Maeno, Phys. Rev. Lett. {\bf 84} 2662 (2000).
\bibitem{rice95} T.M. Rice and M. Sigrist, J. Phys.: Condens. Matter {\bf 7},
L643-L648 (1995).
\bibitem{ishida98} K. Ishida {\it et al.}, Nature {\bf 396},
 658 (1998).
\bibitem{duffy00} J. A. Duffy {\it et al.}, Phys. Rev. Lett. {\bf 85}
 5412 (2000).
\bibitem{nishizaki00} S. NishiZaki, Y. Maeno and Z. Mao, J. Phys. Japan
{\bf 69}, 336 (2000).
\bibitem{bonlade00} I. Bonalde {\it et al.}, Phys. Rev. Lett.
{\bf 85}, 4775 (2000).
\bibitem{izawa00} K. Izawa {\it et al.}, Phys. Rev. Lett. {\bf 86},
2653 (2001).
\bibitem{luke98} G.M. Luke {\it et al.} Nature {\bf 394} 558 (1998).
 \bibitem{volovik85} G.E. Volovik and L.P. Gorkov,
Zh. eksp. teor. Fiz. {\bf 88}, 1412 (1985) [ Sov. Phys.
JETP, {\bf 61}, 843 (1985)].
\bibitem{ozaki86} M. Ozaki, K. Machida and T. Ohmi,
Prog. Theor. Phys. {\bf 75}, 422 (1986).
\bibitem{sigrist87} M. Sigrist and T. M. Rice, Z. Phys. B {\bf 68},
9 (1987).
\bibitem{ozaki89} M. Ozaki and K. Machida, Phys. Rev. B {\bf 39},
4145 (1989).
\bibitem{annett90} J.F. Annett, Adv. Phys. {\bf 39},  83 (1990).
\bibitem{sigrist91} M. Sigrist and K. Ueda, Rev. Mod. Phys.
{\bf 63}, 239 (1991).
\bibitem{graf00} M.J. Graf and A.V. Balatsky, Phys. Rev. B {\bf 62},
9697 (2000).
\bibitem{won00} H. Won and K. Maki, Europhys. Lett. {\bf 52},  427-433 (2000).
\bibitem{dahm00} T. Dahm, H. Won, and K. Maki, cond-mat/0006301.
\bibitem{eremin01} I. Eremin, D. Manske, C. Koas and K.H.
Bennemann, cond-mat/01020774.
\bibitem{manske02} D. Manske, I. Eremin and K.H. Bennemann, in {\it
New Trends in Superconductivity}, J.F. Annett and S. Kruchinin
(eds.) 293-305, (Kluwer, 2002).
\bibitem{agterberg97} D. F. Agterberg, T.M. Rice and M. Sigrist, Phys. Rev.
Lett. {\bf 73} 3374 (1997).
\bibitem{zhitomirsky01} M.E. Zhitomirsky and T.M. Rice,
 Phys. Rev. Lett. {\bf 87},
 057001 (2001).
\bibitem{annett01} J.F. Annett  G. Litak, B.L. Gyorffy and
K.I. Wysoki\'nski, Phys. Rev. B {\bf 66}, 134514 (2002).
\bibitem{hasegawa00} Y. Hasegawa, K. Machida and M. Ozaki, J. Phys. Japan
{\bf 69}, 336 (2000).
\bibitem{mackenzie96} A.P. Mackenzie {\it et al.}, Phys. Rev. Lett.
{\bf 76}, 3786 (1996). Note the corrected Fermi surface parameters
in: Y. Maeno {\it et al.}, J. Phys. Soc. Japan {\bf 66}, 1405
(1997).
\bibitem{szotek} Z. Szotek, B.L. Gyorffy, W.M. Temmerman, O.K. Andersen, O. Jepsen, J
Phys. Condens. Mat. {\bf 13}, 8625
(2001).
\bibitem{miyake} K. Miyake and D. Narikiyo, Phys. Rev. Lett. {\bf 83} 1423
(1999).

\bibitem{mar99} A.M. Martin, G. Litak, B.L. Gy\"{o}rffy, J.F. Annett and
 K.I. Wysoki\'nski, Phys. Rev. B 60 7523 (1999). 
\bibitem{lit00} G. Litak, J.F. Annett, B.L. Gy\"{o}rffy, Acta
Phys. Pol.  A {\bf 97}, 249 (2000). 
\bibitem{lit01} G. Litak, J.F. Annett, B.L. Gy\"{o}rffy, in {\it
Open Problems in Strongly Correlated Electron Systems} Ed. J. Bonca {\em et al.} (Kluwer
Academic Publishers NATO Science Series, Dordrecht 2001) pp. 425--427. 
\bibitem{lit02c} G.
Litak, Phys. Stat. Sol. B 229, 1427 (2002). 
\bibitem{agt99} D.F. Agterberg, Phys. Rev. B {\bf 60}
R749 (1999). 


\bibitem{han75} E. R. Hansen, A Table of Series and Products,
(Prentice Hall, Inc., Englewood Cliffs, N.J. 1975).

\bibitem{micnas} R.Micnas {\it et al.},
 Rev. Mod. Phys. {\bf 62}, 113 (1991).
\bibitem{zawadowski}  D.L. Cox and A. Zawadowski,
{\it Exotic Kondo Effects in Metals}, (Taylor and Francis, London
1999).
\bibitem{lit02} G. Litak, J.F. Annett, B.L. Gyorffy and K.I. Wysoki\'nski,
in {\it New Trends in Superconductivity} Eds. J.F. Annett and S.
Kruchinin (Kluwer
 Academic
Publishers, Dordrecht 2002)  pp. 307--316.
\bibitem{wys02} K.I. Wysoki\'nski, G. Litak, J.F. Annett, B.L. Gy\"{o}rffy,
Phys. Stat. Sol. B {\bf 236} 325 (2003).

\end{thebibliography}
\end{document}